\documentclass[manuscript]{aastex}

\usepackage[dvips]{color}
\definecolor{red}{rgb}{1, 0, 0}

\definecolor{blue}{rgb}{0, 0, 1}

\slugcomment{Draft Version \today}

\shorttitle{The M31 Near-Infrared Period-Luminosity Relation and its non-linearity for $\delta$ Cep Variables with $0.5 \leq \log(P) \leq 1.7$}
\shortauthors{Kodric et al.}

\begin{document}

\title{The M31 Near-Infrared Period-Luminosity Relation  and its non-linearity for $\delta$ Cep Variables with $0.5 \leq \log(P) \leq 1.7$}

\author{Mihael Kodric\altaffilmark{1,2}, Arno
  Riffeser\altaffilmark{1,2}, Stella Seitz\altaffilmark{1,2}, Jan
  Snigula\altaffilmark{2,1}, Ulrich Hopp\altaffilmark{1,2}, Chien-Hsiu
  Lee\altaffilmark{2,1}, Claus Goessl\altaffilmark{1,2}, Johannes
  Koppenhoefer\altaffilmark{2,1}, Ralf Bender\altaffilmark{2,1}, Wolfgang Gieren\altaffilmark{3,4}}

\email{kodric@usm.lmu.de}

\received{May 20, 2014}
\accepted{Nov 19, 2014}

\altaffiltext{1}{University Observatory Munich, Scheinerstrasse 1, 81679 Munich, Germany}
\altaffiltext{2}{Max Planck Institute for Extraterrestrial Physics, Giessenbachstrasse, 85748 Garching, Germany}
\altaffiltext{3}{Universidad de Concepci\`{o}n, Departamento de
  Astronomia, Casilla 160-C, Concepci\`{o}n, Chile}
\altaffiltext{4}{Millenium Institute of Astrophysics, Av. Vicu{\~n}a Mackenna 4860,
   Santiago, Chile} 

\begin{abstract}

We present the largest M31 near-infrared (F110W (close to J band),
F160W (H band)) Cepheid sample so far. The sample consists of
371 Cepheids with photometry obtained from the HST PHAT program. The
sample of 319 fundamental mode Cepheids, 16 first overtone Cepheids
and 36 type II Cepheids, was identified using the median absolute
deviation (MAD) outlier rejection method we develop here. This method
does not rely on priors and allows us to obtain this clean Cepheid
sample without rejecting a large fraction of Cepheids. The obtained
Period-Luminosity relations (PLRs) have a very small dispersion,
i.e. 0.155 mag in F160W, despite using random phased
observations. This remarkably small dispersion allows us to determine
that the PLRs are significantly better described by a broken slope at
ten days than a linear slope. The use of our sample as an anchor to
determine the Hubble constant gives a $3.2\%$ larger Hubble constant
compared to the \citet{Riess} sample.

\end{abstract}

\keywords{catalogs -- cosmology: distance scale -- galaxies: individual(M31) -- Local Group -- stars: variables: Cepheids}

\section{Introduction}

The Cepheid period-luminosity relation (PLR) remains an important rung
of the cosmic distance ladder, and is an integral means of
establishing the Hubble constant (\citet{2006ApJ...653..843S},
\citet{2010ARA&A..48..673F} and \citet{Efstathiou} (E14)).

Apart from using Galactic Cepheids to establish the PLR calibration,
another place that is usually used for this calibration is the Large
Magellanic Cloud (LMC). Extensive studies have been conducted to study
the variable stars content in the LMC with the OGLE project probably
being the most extensive \citep{1999AcA....49..223U}. Cepheids in
the Andromeda galaxy (M31) belong to the closest spiral galaxy
exhibiting near-solar abundances. Observations of these Cepheids are
particularly important since the impact of metallicity on the PLR is
actively debated (e.g., \citet{2010ARA&A..48..673F},
\citet{2011ApJ...741L..36M}). Furthermore a debate continues
concerning the existence of a broken PLR slope \citep{Sandage}. Both
these effects may impact the establishment of the Hubble constant and
the cosmic distance scale. The difficulty with M31 is its crowding
(overlapping point spread functions (PSFs)) and blending caused
by the high inclination. In order to obtain a representative sample of
the whole galaxy, the large angular size makes wide field CCDs
necessary.  For a recent summary of ground based Cepheid observations
in M31 see \citet{kodric2013} (hereafter K13).

\citet{2008A&A...477..621N} applying statistical tests such as an
F-test find a broken slope at 10 days in the BVIcJH bands but a linear
relation in the Ks band and the Wesenheit
functions. \citet{2013ApJ...764...84I} on the other hand find that
their Magellanic Cloud Period-Wesenheit relations are
linear. \citet{2013MNRAS.431.2278G} observe non linear relations in
the VI bands and that the Wesenheit function behaves exponentially.

Near-infrared photometry has the advantage that the extinction is low
\citep{1982ApJ...257L..33M} and that the amplitudes of the
Cepheids are usually smaller than in the optical
\citep{1991PASP..103..933M}. The increase in the dispersion of the PLR
caused by random phased observations is minimized for small
amplitudes. Hubble space telescope (HST) observations in the
near-infrared allow for very precise PLRs with small dispersion as
shown recently by \citet{Riess} (hereafter R12). Nevertheless there
are also Cepheids with near-infrared amplitudes of around 0.5
mag. These Cepheids increase the dispersion if random phased
observations are used. An outlier rejection mitigates this
problem. The optimal solution is to use mean phase observations or
perform a phase correction. HST observations also help with the
problem of crowding. The ground-based observations are in this
case only used to identify the position of the Cepheid and to obtain
the period of the Cepheid.

In this paper we follow this approach and combine ground based
observations with near-infrared HST observations. As a Cepheid sample
we use the 2009 Cepheids published in K13. The HST observations are
from the PHAT survey of M31 \citep{PHAT}. The PHAT data cover roughly
a third of the disk of M31 in 6 filters (F275W, F336W, F475W, F814W,
F110W and F160W) with two orbits per pointing. The relative difference
to R12 is that we included all three years of PHAT observations that
are now available and that our Cepheid sample (with up to 180
photometric epochs) is published and available in CDS. The
\citet{2012Ap&SS.341...57F} sample (up to 50 epochs) which is used in
R12 is larger but not yet publicly available. As discussed in E14 the
R12 outlier rejection procedure can lead to underestimated errors in
the PLR parameters. We develop an outlier rejection procedure that is
similar to the one proposed by E14, but more robust (i.e. the
convergence is less susceptible to starting parameters). Another
change compared to R12 is that we develop a sophisticated pipeline
that uses difference images to identify the correct source in the PHAT
data instead of relying on information from the UV filters when the
source identification is unclear. The reason is that there is no UV
information for each Cepheid, while good HST difference images are
available for almost all Cepheids.

The paper is structured as follows: Section \ref{datareduction}
discusses the data reduction and how to identify the correct source in
the PHAT data. Section \ref{section_outlier} describes our
outlier rejection procedure. The Period-Luminosity relations are
discussed in section \ref{chapter_PLRs}. The impact of the
improved PLRs on the Hubble constant is examined in section
\ref{chapter_H0} followed by the conclusions in section
\ref{conclusion}.

\section{Data analysis\label{datareduction}}

The goal is to obtain near-infrared photometry of the Cepheid sample
published in K13. The K13 Cepheid sample
contains 2009 Cepheids obtained during the first year (up to 180 epochs) of PS1
PAndromeda observations \citep{shermanPAndromeda}.  The sample
consists of 1440 fundamental mode (FM) Cepheids, 126 first overtone
(FO) and 147 type II Cepheids. For 296 Cepheids the type of Cepheid
could not be assigned. The Cepheid type was automatically assigned in a three
dimensional space of period, amplitude ratio and phase difference, where
the last two parameters were obtained from Fourier decomposition of
the light curve. In order to obtain near-infrared photometry we match
this data set with the PHAT data \citep{PHAT}.

We obtained the PHAT data in November 2013 from the MAST archive. At
that time photometry was not available for all bricks. Therefore we
ran DOLPHOT \citep{DOLPHOT} on all data with the same parameters that
were used on the already available photometry in the MAST
archive. Additionally we put artificial stars into the images and
tested the impact of crowding on the photometry of the Cepheids.
For each Cepheid we put an artificial star of the magnitude of the
Cepheid in proximity to the Cepheid. We do this iteratively 10000
times in order to estimate the impact the environment of the Cepheid
has on the photometry. As expected for crowding a very close source 
to the artificial Cepheid causes the recovered magnitude of the
artificial star to be brighter. With this procedure it is possible to
test the effect of overlapping PSFs, i.e. crowding, but not the impact
of blending. Above a certain distance between fake source and the
corresponding closest source the recovered magnitude should match
the magnitude of the source that was put into the image.

\begin{figure}
\epsscale{1.0}
\plotone{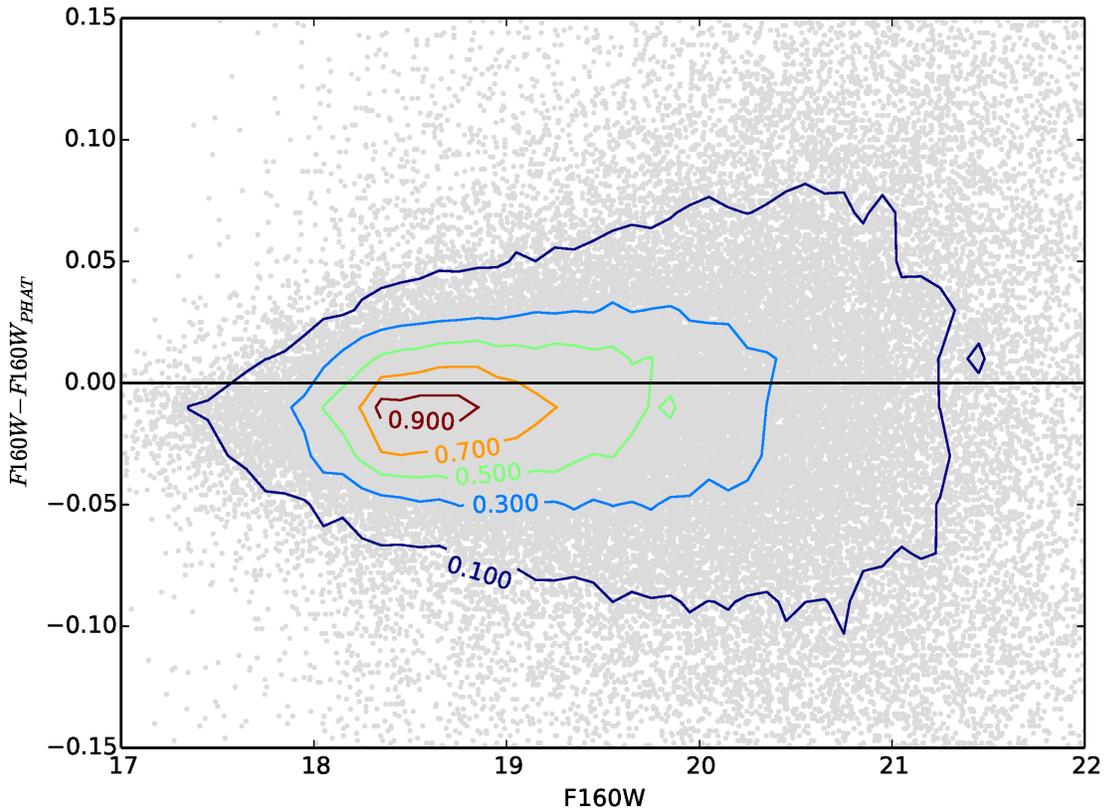}
\caption{Comparison between our F160W photometry of the published
    PHAT data and the published PHAT photometry catalogs (Brick 01,
    Field 09). For this comparison we use the same DOLPHOT parameters
  that were used in the PHAT catalog. The distribution of the points
  is illustrated with contour lines of the two-dimensional
  histogram. The lines show the contours where the histogram falls to
  90, 70, 50, 30 and 10 percent of the peak density. The small
    difference is due to the fact that we use the Anderson PSFs that
    take into account the spatial variation over the field of
    view. The median difference between the standard DOLPHOT PSF and
    the Anderson PSF is $-0.015$ mag which is also the offset we see
    in the comparison shown here. \label{fig_PHAT_color}}
\end{figure}

\begin{figure}
\epsscale{0.5}
\plotone{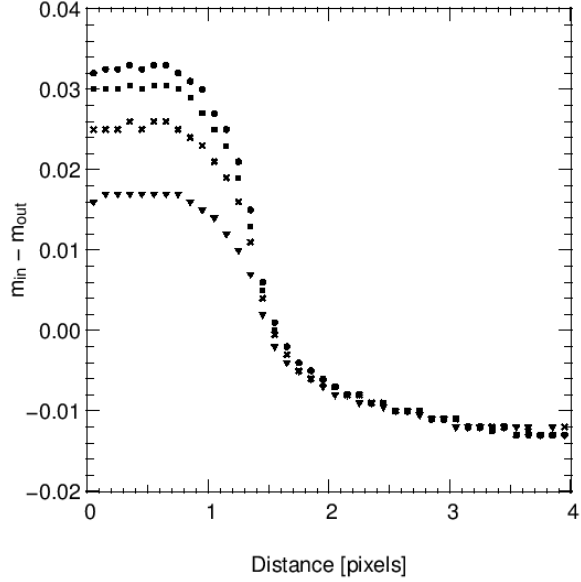}
\caption{Impact of crowding on our Cepheid sample depending on the
  distance of the closest source. For each Cepheid a fake source with
  the same magnitude as the Cepheid is put in the proximity of the
  Cepheid. This is done iteratively 10000 times for each Cepheid. The
  difference in magnitude of the fake source ($m_{\mathrm{in}}$) and
  the recovered source ($m_{\mathrm{out}}$) is a measure of the impact
  of the Cepheids environment on the crowding. This median difference
  for all iterations of all Cepheids is shown for different distance
  bins of the closest source to the fake
  source. $m_{\mathrm{in}}-m_{\mathrm{out}}$ should be zero for large
  distances, but due to the background determination it is $\approx
  -0.01~\mathrm{mag}$. This behavior gets worse if the standard
  background determination parameter is used instead of the parameter
  for highly crowded fields that is used here. Crowding is only
    relevant for sources that have the closest source closer than 1.5
    pixels. For separations closer than 1 pixel the pixel quantization
    causes a plateau. Even then the magnitude typically changes only by
    $0.035~\mathrm{mag}$. For these very close separations DOLPHOT
    might not recover the fake source, which means that there is
    blending (which is not examined here). The crowding does also
    depend on the magnitude difference between the fake source and the
    closest source. The triangles show the crowding for magnitude
    differences of $3$ mag or larger (i.e. the fake source is at least
    $3$ mag brighter that the closest source), the crosses for $2$ mag
    or larger, the squares for $1$ mag or larger and the points for
    all magnitude differences (including the cases where the fake
    source is fainter than the closest source). \label{fig_crowding}}
\end{figure}

\begin{figure}
\epsscale{1.0}
\plotone{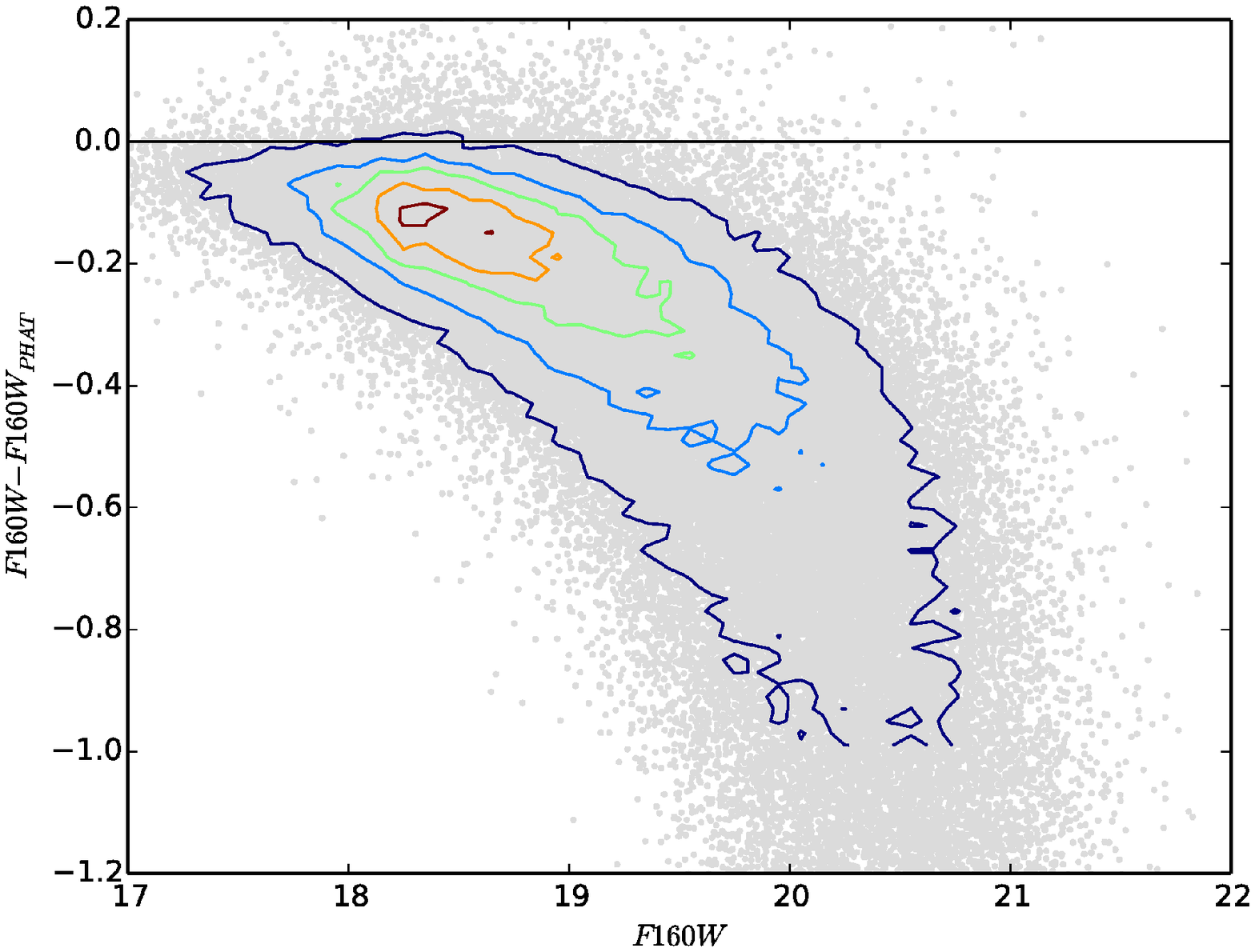}
\caption{Same as Fig. \ref{fig_PHAT_color} but with a DOLPHOT sky
  fitting parameter that is recommended for highly crowded fields. Due
  to the change in the background determination method our photometry
  is not consistent any more to the published PHAT catalog. The
    sky fitting affects the faint stars more than the bright
    stars. This trend might affect the slope of the PLR but indeed the
    photometry of the Cepheids only changes slightly due to their
    brightness. The results are not significantly affected by the
    change in sky fitting parameter.
  \label{fig_PHAT_color_new}}
\end{figure}

For a field close to the center of M31 (Brick 01, Field 09) we compare
the photometry of the already published PHAT catalog with the
photometry we obtain when we use the same DOLPHOT parameters as in the
PHAT catalog (the catalog also includes the parameter files). As can
be seen in Fig. \ref{fig_PHAT_color} our photometry matches that of
the published PHAT catalog in this field.  The small offset can be
attributed to the fact that we make use of the improved Anderson PSFs
\citep{andersonpsf} in our photometry. The Anderson PSFs take into
account the spatial variation over the field of view. But when we
investigate the crowding of the Cepheids using these DOLPHOT
parameters we observe a strange behavior. The recovered magnitude of
the fake star is fainter if there is no source close by,
i.e. $m_{\mathrm{in}}-m_{\mathrm{out}} < 0~\mathrm{mag}$. This effect
is of the order of $m_{\mathrm{in}}-m_{\mathrm{out}} \approx 0.04$ mag
for a closest source separation of 4 pixels (crowding becomes relevant
for separations closer than 1.5 pixels). This problem seems to be
caused by the background determination, since the flux of the
artificial star can only be attributed to the background, due to the
lack of other sources nearby. A change of the sky fitting parameter of
DOLPHOT from the default parameter that is used by PHAT, to the one
recommended for highly crowded fields alleviates the problem,
i.e. $m_{\mathrm{in}}-m_{\mathrm{out}} \approx 0.01$ mag. Using this
parameter we can see in Fig. \ref{fig_crowding} that crowding is not
present for closest neighbor distances larger than 1.5 pixels and that
the crowding typically changes the magnitude of the Cepheid by no more
than $\approx 0.035$ mag. Of course this is only statistically true
and the real change of the magnitude can be higher and is also
dependent on the magnitude of the source that is close to the Cepheid.
Changing the background determination parameter also changes the
photometry. The comparison to the PHAT catalog can be seen in
Fig. \ref{fig_PHAT_color_new} . The photometry of the Cepheids is only
slightly affected by the change of the background parameter. The
results are also not significantly changed by the different sky
fitting. The impact of crowding on the photometry of the Cepheids is
also very small and we therefore use the complete Cepheid
sample. Although our results do not change significantly due to
crowding, for the interested reader the appendix includes all results
without using the Cepheids that have sources closer than 1.5 pixels.

We developed a pipeline that identifies the Cepheids from the first
year of PAndromeda observations (K13) in the PHAT data.  For each
Cepheid the pipeline astrometrically matches the corresponding PHAT
frames (of all filters) to the PS1 reference frame.  After that step
we create stamp outs (i.e. small images around the Cepheid) from the
aligned PHAT data and the PS1 data. Additionally the pipeline produces
difference images from the PHAT data. The number of epochs for each
Cepheid is highly dependent on its position (i.e. if it is in an
overlap of PHAT bricks). However, due to the observing strategy the optical
PHAT filters have at least two epochs (c.f. Fig. 5 in
\citet{PHAT}). Then the Cepheid is identified automatically from these
PHAT difference images. This rather sophisticated procedure is
necessary due to the fact that it is often unclear which source is the
Cepheid in question, as the HST images resolve the error circle of the
PS1 source into typically multiple sources. To make sure that the
correct source is selected we inspect the result from the pipeline by
eye. This involves checking the PHAT stamp outs and difference frame
stamp outs of each Cepheid for consistency. This means making sure
that the same source is selected in all filters\footnote{Note that we
  do allow the pipeline to find different coordinates in each
  filter. This way we obtain another quality check for the
  determination of position from the difference frames.}. The pipeline
works remarkably well and the few times it fails\footnote{Usually it
  fails when there are only two frames available that are taken
  shortly after one another causing the resulting difference frame to
  show small variability.} the information from the WFC3-UVIS frames
helps to identify the correct source visually. R12 use the UVIS
information to select the correct source when there is a close
neighboring source. Although the UVIS data can be very helpful, the
problem with this approach is that there are not always UVIS
observations available and that the UVIS data can be too shallow to
find the source. This is why a source identification based solely on
the UVIS information proved to be inferior to the difference image
method.

We were able to identify 557 Cepheids from the 2009 Cepheids published
in K13 in the PHAT data (all bands)\footnote{The main reason for
  finding no PHAT counterparts is the smaller sky coverage of PHAT
  compared to the PS1 data set. 1515 Cepheids of the 2009 are outside
  the area covered by F160W observations.}. 528 have F110W (close to J
band) measurements and 494 have F160W (H band) measurements. 492
Cepheids have F110W and F160W data. While we use all bands for the
source identification, in this paper we will only discuss the Cepheids
with WFC3-IR data. The obtained magnitudes are random phased. We
perform no phase correction since the PS1 epochs in K13 do not cover
all PHAT epochs. The precision of the periods from just the first year
of PAndromeda observations can be insufficient to determine the
correct phase for some PHAT epochs two years apart from the K13
data. With the full PS1 data set of three years we will be able to
perform phase corrections. In the few cases in which multiple PHAT
measurements are available we therefore use the mean magnitude.

\begin{figure}
\epsscale{1.0}
\plotone{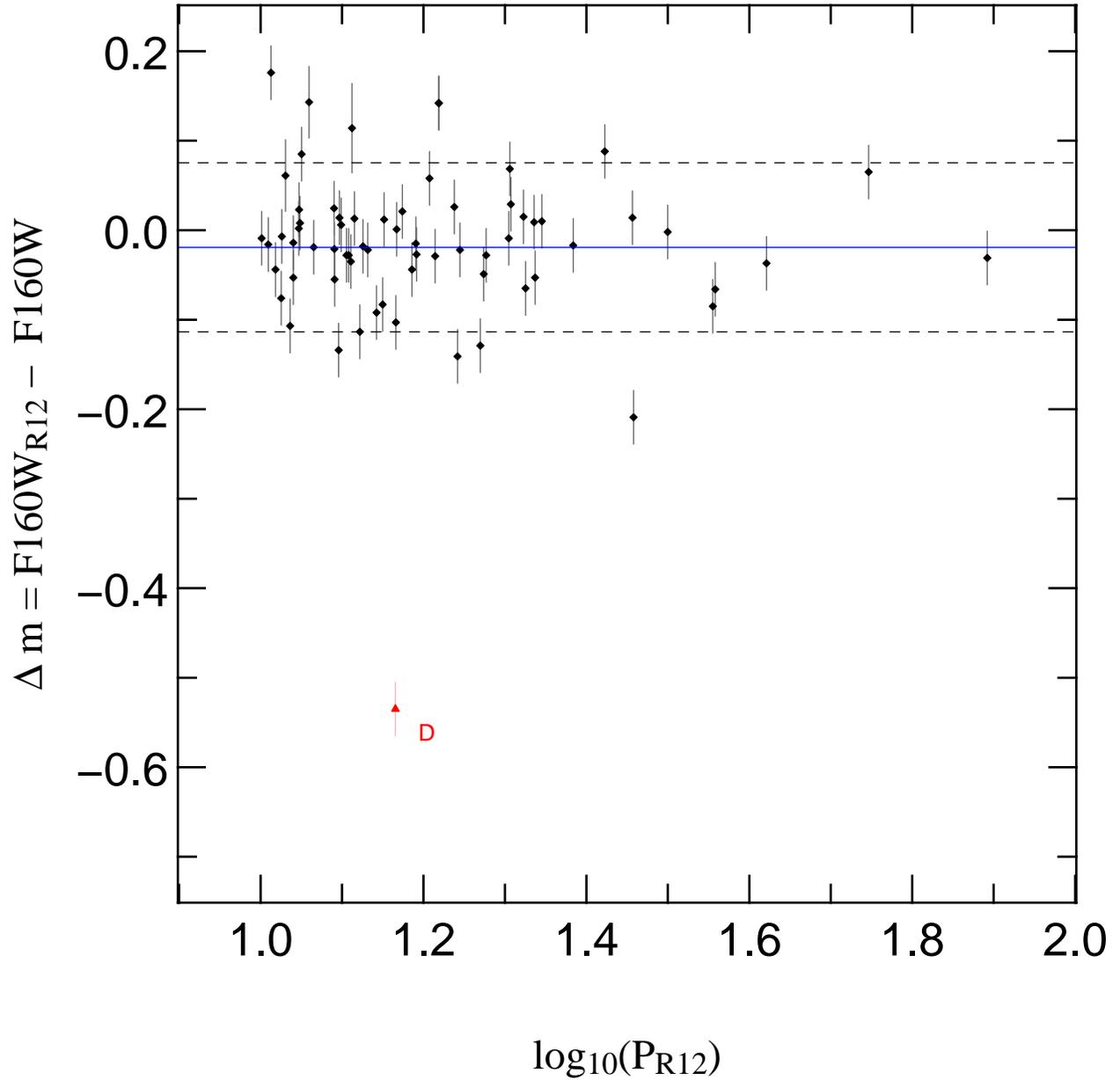}
\caption{Comparison of our photometry with R12. The Cepheid
  marked in red (inverted triangle) with a D is due to a misidentification in
  R12. The mean magnitude difference is $\Delta m =
  -0.019 \pm 0.011$ mag (blue solid line). The standard deviation
  is shown as a black dashed line.  \label{fig_compare_riess_160}}
\end{figure}

\begin{figure}
\epsscale{1.0}
\plotone{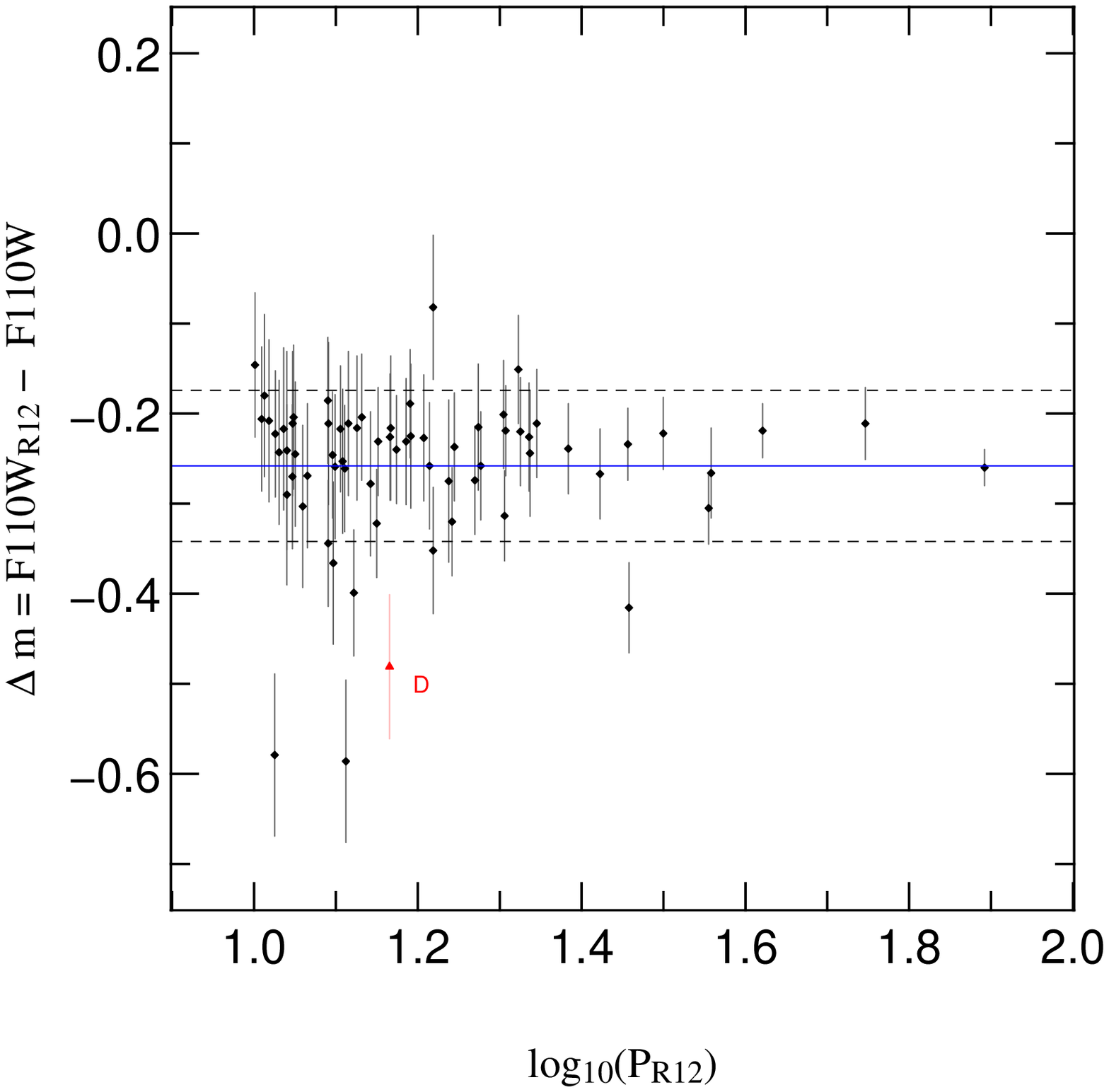}
\caption{Same as Fig. \ref{fig_compare_riess_160} but with a mean
  $\Delta m = -0.258 \pm 0.010$ mag. The two outliers with $\Delta m < -0.5$
  mag are likely due to a close second source that contaminates the aperture
  photometry of R12. We reproduce the large difference in
  photometry when performing  aperture photometry ourselves. The
    offset can be explained by an error in the aperture correction in
  R12.\label{fig_compare_riess_110}}
\end{figure}

To check our photometry we compare it to R12. 51 of the 68 R12
Cepheids are contained in the K13 sample. Cepheid vn.2.2.463 is
present twice in the R12 sample with the same identifier, position,
period and F160W photometry, but with a different F110W photometry. So
there are rather 50 of the 67 R12 Cepheids contained in K13. We run
the remaining 17 Cepheids through our pipeline and include them in our
comparison. We compare the stamp outs provided in R12 to our source
identification and find only one deviation. For Cepheid vn.2.3.69 (PSO
J011.4455+41.9120) our difference frames indicate that the variable
source (Cepheid) is indeed the source next to the one identified in
the R12 stamp out. We marked this Cepheid with a red D in
Fig. \ref{fig_compare_riess_160} and Fig. \ref{fig_compare_riess_110}.
Fig. \ref{fig_compare_riess_160} shows a mean magnitude difference in
F160W photometry of $-0.019 \pm 0.011$
mag. Fig. \ref{fig_compare_riess_110} indicates a mean magnitude
difference of $-0.258 \pm 0.010$ mag in F110W. The two outliers below
$\Delta m < -0.5$ mag have a very close source nearby and the offset
can be explained by the fact that R12 use aperture photometry in
F110W. However that does not explain the offset of approximately a
quarter of a magnitude in F110W. This difference remains approximately
the same if we perform aperture photometry. The reason for this offset
is that R12 used the STScI table for the aperture correction that
gives the ensquared energy fraction vs. the aperture size in pixels
but assumed this to be the encircled energy fraction (A. Riess, private
communication 2014). This explains the offset in F110W.  We
conclude that our photometry matches that of R12.

\section{Outlier rejection\label{section_outlier}}

\begin{figure}
\epsscale{1.0}
\plotone{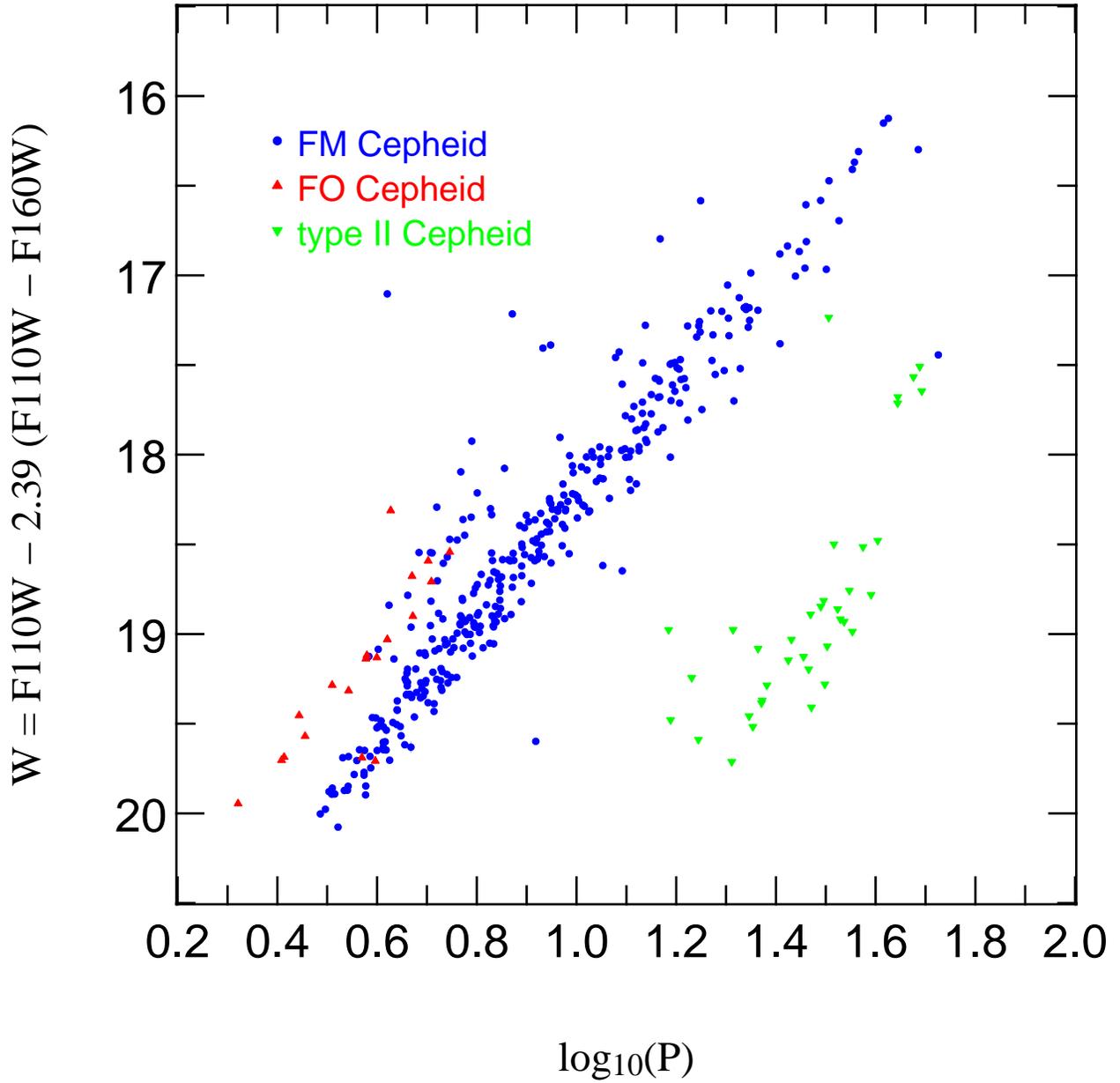}
\caption{Unclipped Wesenheit Period-Luminosity diagram of 413 classified
  Cepheids. The shown classification is from
  K13. The unclassified Cepheids are omitted. The
  diagram shows the necessity of outlier rejection. The different
  reasons for outliers as well as an outlier rejection procedure are
  discussed in Section
  \ref{section_outlier}.\label{fig_Wesenheit_raw}}
\end{figure}

After finding 492 Cepheids with F110W and F160W photometry in the PHAT
data we want to investigate the Period-Luminosity relation (PLR). As a
first step we have to exclude the outliers of our sample that can be
seen in Fig. \ref{fig_Wesenheit_raw}. The Wesenheit magnitude, which
is reddening-free, used in
this figure is defined as:
\begin{equation}
W =  m_{F110W} - R \cdot (m_{F110W}-m_{F160W})
\label{eqn_Wesenheit_1}
\end{equation}
where $R$ can be obtained from \citet[table 6 with $R_V=3.1$]{2011ApJ...737..103S}
\begin{equation}
R = \frac{A_{F110W}}{A_{F110W}-A_{F160W}}=2.39
\label{eqn_Wesenheit_2}
\end{equation}

There are different reasons for outliers in the period-luminosity
relation, namely blending, crowding, extinction, misidentification and
misclassification. In the case of blending multiple sources sit along
the same line of sight. This is the most difficult case to resolve due
to the fact that it needs extensive modeling to do so.
\citet{2007A&A...473..847V} studied the impact of blending on the M31
distance and concluded that blending impacts the M31 distance on a
$\sim 0.1$ mag level which makes it as significant as the impact of
metallicity. Crowding introduces errors in the photometry due to
overlapping point spread functions (PSFs). This is obviously worse in
ground based observations where the PSFs are larger. The Hubble space
telescope (HST) PSF is well determined and stable and as discussed in
the previous section (see Fig. \ref{fig_crowding}), crowding does not
significantly contribute to our photometric errors. Determining the
correct extinction for each Cepheid with spectroscopy is un practical
for Cepheids in M31 due to the long exposure times needed and the
large spatial extent of M31. In our case we have NIR photometry
available for which the extinction is low
\citep{1982ApJ...257L..33M}. Another way to get a handle on extinction
is to use Wesenheit magnitudes $W$ that are independent of reddening.

The simplest cause for an outlier is misidentification, i.e. selecting
the wrong source when matching two samples. Due to the method of
identifying the PS1 Cepheid in PHAT from difference images this kind
of mismatch should not be present in our sample. A misclassification
of the Cepheid type (fundamental mode (FM), first overtone (FO) and
type II) or the identification of a different kind of variable as
  a Cepheid can also lead to an outlier in the PLR.

The Cepheid type determined by K13 is biased by blending and
crowding. Separating FM and FO Cepheids in M31 using ground-based
  observations is difficult. Ideally, the type would be determined
with near infrared light curves. For larger wavelengths the scatter in
the PLR is smaller because the temperature sensitivity on the surface
brightness is smaller for longer wavelengths
\citep{2012ApJ...744..132M}. Even in HST data a Cepheid that is
clearly FO in the F160W PLR scatters into the FM in the F814W
PLR\footnote{We see this behavior in the data of the optical bands,
  which we do not discuss in this paper.}. For this reason we exclude
all unclassified Cepheids\footnote{Cepheids where the type could not
  be determined.}  from K13 from our sample. This leaves us with 447
Cepheids in F110W and 415 Cepheids in F160W. 413 Cepheids have
photometry in both bands simultaneously.

The typical photometric errors we get from DOLPHOT are 0.003
mag. These are very small and do not account for the dispersion of the
PLR. The photometric errors are only one aspect that contributes to
the dispersion. Extinction and the inherent width of the PLR due to
the temperature dependence of the instability strip
\citep{1958ApJ...127..513S} are other aspects. In the case of the
Wesenheit PLR, different extinction laws for each Cepheid would change
$R$ (Equation \ref{eqn_Wesenheit_2}) and therefore increase the
scatter in the PLR.  The photometric errors in R12 are also very small
and as mentioned in E14, \citet{Riess2011} add 0.21 mag in quadrature
to the magnitude errors.  An ordinary clipping routine without priors
or rescaling of the magnitude errors performs very poorly. Introducing
priors and rescaling the errors works, but that either usually clips a
large fraction of the data or the outlier rejection is
unsatisfactory. Testing this method we found no working compromise
between clipping away way too much or almost nothing. The problem of
outlier clipping and potential implications on the PLR-biases has been
recently investigated in detail by E14. As pointed out by E14 that
approach possibly underestimates the errors of the PLR. Additionally
the combination of priors and strong clipping would prevent a study of
the broken slope in our data as was done by \citet{Sandage} for their
BVI data.  On the other hand stricter outlier rejection leads to less
blending in the crowded central region of a galaxy
\citep{2013ApJ...777...79M}

We therefore develop a simple outlier rejection method that does
not rely on any prior. In the first iteration of the algorithm we
assign all measurements the same error and perform a linear fit. The
error we assign in the first iteration is the average magnitude
error. This ensures (empirically) that at least one Cepheid is above
the clipping threshold\footnote{The dispersion of the initial fit can
  be so large that nothing would be clipped if this large dispersion
  would be chosen as the error.}.  After excluding the largest outlier
to that fit we calculate the dispersion. For the next step we set the
median of the absolute regression residuals (median absolute
deviation; MAD) as the magnitude error. After the fit the worst
outlier over a threshold of $\kappa$ times the MAD is rejected and the
new MAD is calculated. This is repeated until the procedure
converges. This is a slightly modified $\kappa$ - $\sigma$ clipping
with the MAD for each magnitude error. Another difference to a typical
$\kappa$ - $\sigma$ clipping is that only the worst outlier is clipped
in one iteration step. A normal $\kappa$ - $\sigma$ clipping without a
prior to the slope of the PLR can be heavily influenced by even a few
outliers.  These outliers could influence the PLR fit in a way that
the slope is somewhere between the real PLR and the outliers. The
normal $\kappa$ - $\sigma$ clipping would than clip both from the
outliers and the real PLR. Clipping only one outlier in one iteration
step ensures that an initially wrong PLR fit gradually converges to
the genuine PLR and does not clip non outliers on the way. The reason
for using the MAD instead of the dispersion is that in this way it is
possible to clip Cepheids with a misclassified type, or spurious, or
odd (e.g., Polaris-like) Cepheids. (see Fig. \ref{fig_Wesenheit_raw}).
 
\begin{figure}
\epsscale{0.75}
\plotone{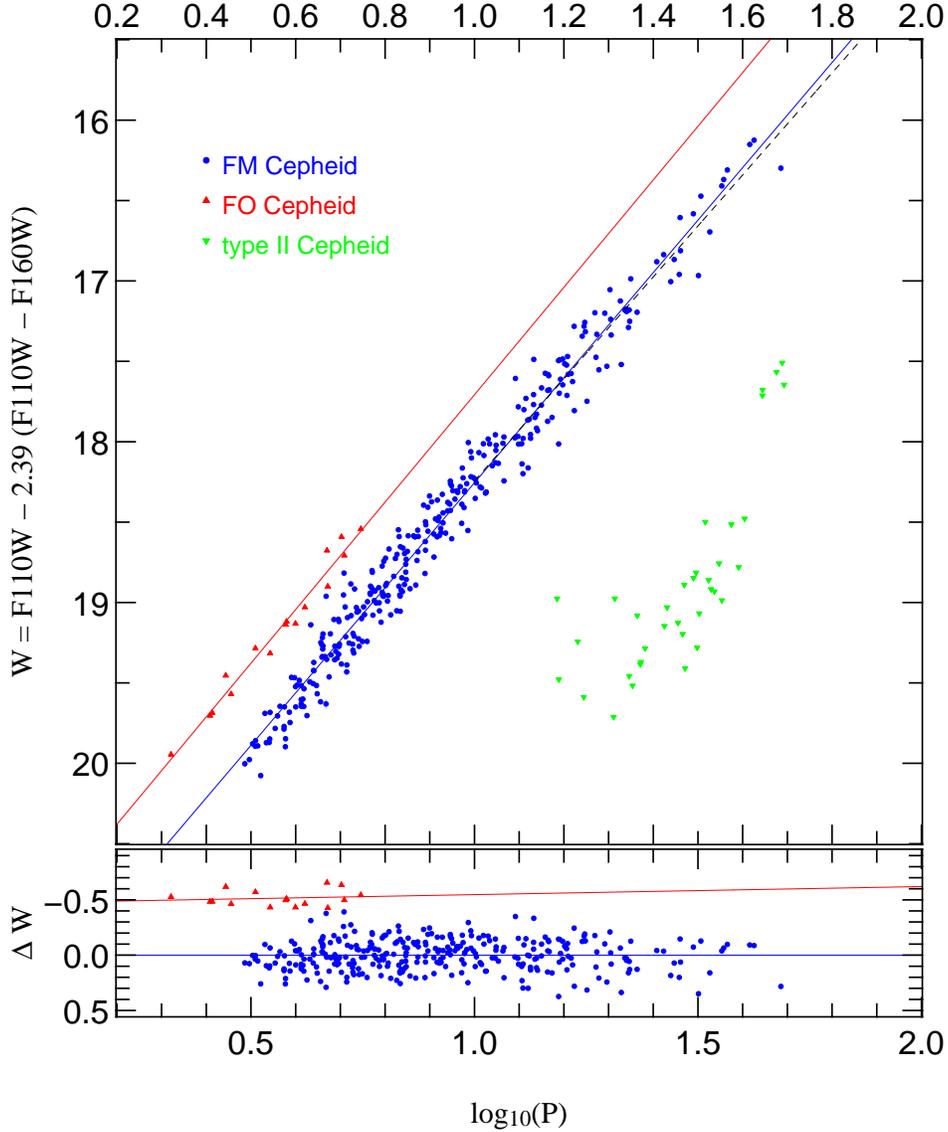}
\caption{Wesenheit Period-Luminosity relation clipped with the MAD
  method. The fit parameters to the FM (319 Cepheids, blue solid line)
  and FO (16 Cepheids, red solid line) PLRs are given in Table
  \ref{table_PLRs} ($\#7$ for the FM and $\#9$ for the FO). The 36
  type II Cepheids apparently show no linear relationship (see Section
  \ref{chapter_PLRs}). The $\log(P)>1$ FM PLR is shown as a black
  dashed line ($\#8$ in Table \ref{table_PLRs}). The photometric
  errors (0.009 mag on average) are smaller than the
  symbols. The bottom panel shows the residuals relative to the
    FM fit.}\label{fig_PLR_Wesenheit}
\end{figure}

\begin{figure}
\epsscale{1.0}
\plotone{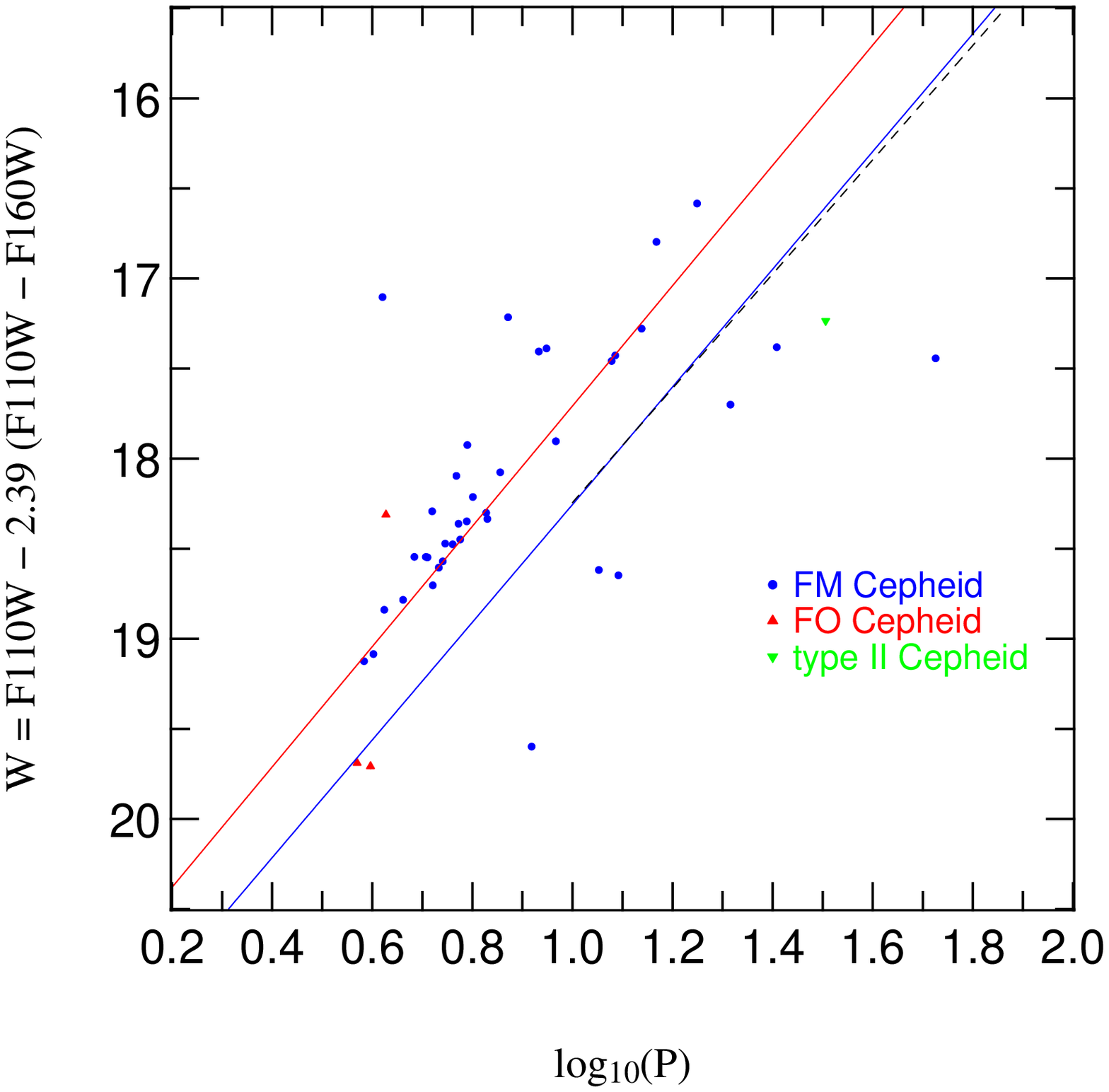}
\caption{Clipped outliers by the MAD method. The PLR
  relations are the same as in Fig. \ref{fig_PLR_Wesenheit}. Most of
  the outliers are most likely due to a misclassified Cepheid
  type. \label{fig_PLR_Wesenheit_outliers}}
\end{figure}

We perform the outlier rejection in the Wesenheit PLR. As a
consequence this means that we need both F110W and F160W photometry
simultaneously and therefore our sample will consist of 413 Cepheids
(FM, FO and type II Cepheids) before the clipping is performed. The
main reason for using the Wesenheit function is to minimize the bias caused by
extinction and to have a homogenous sample in both F110W and F160W.
Clipping in each filter separately could lead to a Cepheid being
rejected in one filter but not in the other. Our $\kappa=4$ clipped
Wesenheit PLR can be seen in Fig. \ref{fig_PLR_Wesenheit}, while the
clipped outliers can be seen in Fig. \ref{fig_PLR_Wesenheit_outliers}.

\begin{figure}
\epsscale{0.75}
\plotone{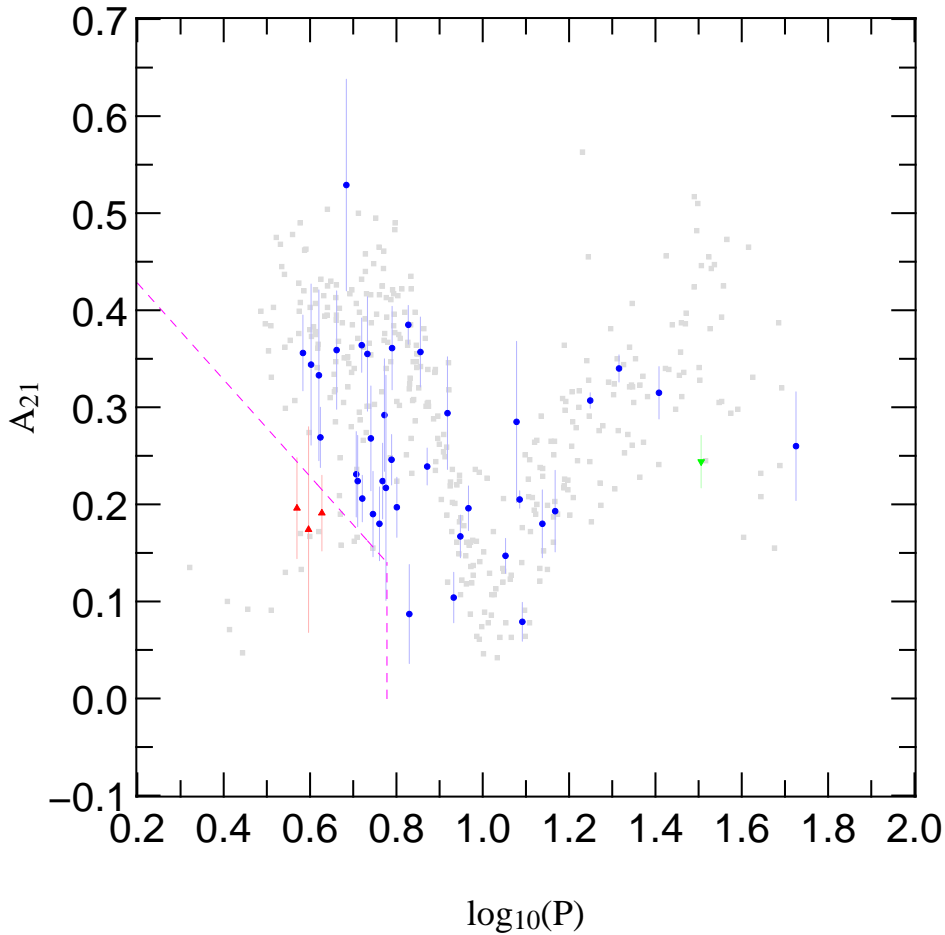}
\caption{Amplitude ratio diagram. The amplitude ratio $A_{21}$ is
  determined from a Fourier decomposition in the $r_{P1}$ band
  (K13). All Cepheids that are not clipped
  (c.f. Fig. \ref{fig_PLR_Wesenheit}) are plotted as gray squares. For
  better visibility the errors for those Cepheids are omitted. The
  clipped Cepheids (c.f. Fig. \ref{fig_PLR_Wesenheit_outliers}) are
  shown as blue circles (FM), red triangles (FO) and inverted green
  triangles (type II). The dashed magenta lines define the boundary
  that was used in K13 to define the parameter space of FO
  Cepheids. As already discussed in K13 there is a transition region
  between FM and FO Cepheids and this is most likely the reason that
  most of the FM outliers that reside on the FO PLR at $\log(P) \sim
  0.75$ (c.f. Fig. \ref{fig_PLR_Wesenheit_outliers}) are misclassified
  as FM and are rather FO. \label{fig_A21}}
\end{figure}

The number of clipped Cepheids is 42 (38 FM, 3 FO and 1 type II),
$\sim$ 10\% of the sample. As can be seen most of our outliers are too
bright in respect to the best fit PLR, which points to
misclassification or blending. About half of the clipped FM Cepheids
rejected reside on the FO PLR. Most of the outliers at $0.55 \lesssim
\log(P) \lesssim 0.85$ are most likely misclassified as FM instead of
being classified as FO. Indeed a lot of them are in a region of the
amplitude ratio ($A_{21}$) diagram (Fig. \ref{fig_A21}) populated by
both FO and FM Cepheids, which makes them difficult to classify. This
is especially true when the light curves, as in our case, are
determined from ground based observations in optical bands. Crowding
and blending will influence $A_{21}$ which contributes to the
misclassification. Blending will decrease amplitudes and the influence
of crowding depends on the magnitude difference of the two sources
(c.f. Fig. \ref{fig_crowding}). Extinction does not influence the type
classification since the classification in K13 only uses the Fourier
parameters of first and second order and the extinction only changes
the zeroth order (i.e. the mean magnitude).  But the greatest
contributing factor for the misclassification will be that FO Cepheids
populate more than the region characterized in the amplitude ratio
diagram in Fig. \ref{fig_A21}. To resolve this issue we would need
spectroscopy or light curves in the near infrared \citep[e.g.,
  see][]{2009MNRAS.396.2194B}. The two clipped sources with the
largest periods are also in a transition region between FM and type II
in the phase difference diagram (see right panel Fig. 9, K13) and
could therefore also be misclassified.

\begin{figure}
\epsscale{0.75}
\plotone{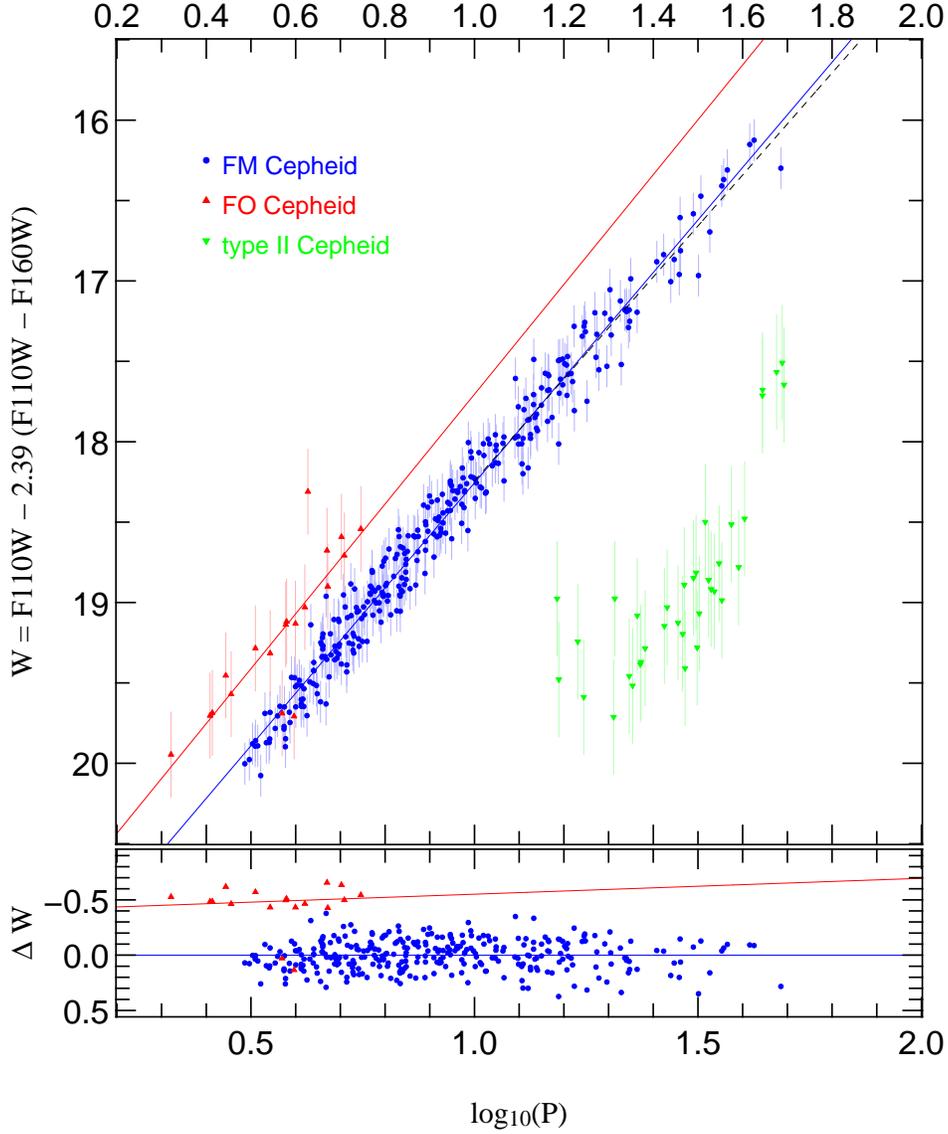}
\caption{Wesenheit Period-Luminosity relation clipped with the
  E14 method. The fit parameters to the FM (blue solid
  line) and FO (red solid line) PLRs are given in Table
  \ref{table_PLRs} ($\#10$ for the FM and $\#12$ for the FO). The
  $\log(P)>1$ FM PLR is shown as a black dashed line ($\#11$ in Table
  \ref{table_PLRs}).  Same as in Fig. \ref{fig_PLR_Wesenheit}
  the type II Cepheids show no linear relationship (see Section
  \ref{chapter_PLRs}). The errors shown
  here are $\sigma = \sqrt{\sigma_{phot}^2 + \sigma_{int}^2}$, where
  $\sigma_{phot}$ is the photometric error which is very small
  compared to the internal scatter $\sigma_{int}$.  \label{fig_Wesenheit_Efstathiou}}
\end{figure}

\begin{figure}
\epsscale{1.0}
\plotone{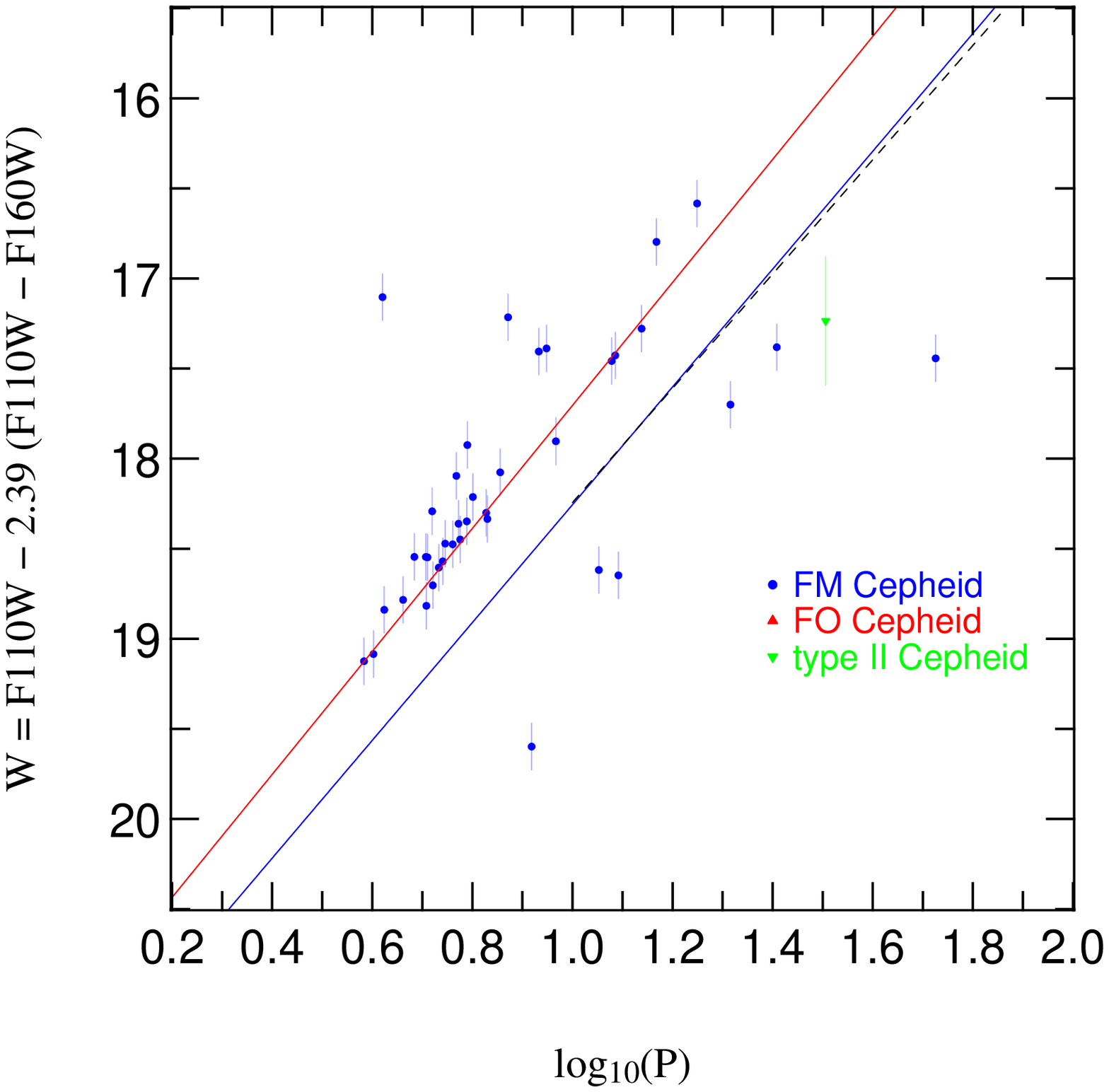}
\caption{Clipped outliers by the E14 method. The PLR
  relations are the same as in
  Fig. \ref{fig_Wesenheit_Efstathiou}. None of the FO Cepheids are
  clipped (3 for the MAD method) and one additional Cepheid is clipped in comparison to
  the MAD clipping method (c.f. Fig. \ref{fig_PLR_Wesenheit_outliers}).
  \label{fig_Wesenheit_Efstathiou_outliers}}
\end{figure}

E14 introduces an internal scatter $\sigma_{int}$ to the
$\chi^2$ minimization in order to obtain a $\chi^2$ of unity:
\begin{equation}
\chi^2 =
\sum\limits_{i}\frac{(m_{W,i}-m_W^P)^2}{(\sigma_{phot,i}^2+\sigma_{int}^2)}
\label{eqn_Efstathiou}
\end{equation}
The clipping is performed iteratively until
convergence. Fig. \ref{fig_Wesenheit_Efstathiou} shows the clipped
Wesenheit PLR if clipped with the E14
method. Fig. \ref{fig_Wesenheit_Efstathiou_outliers} shows the
corresponding outliers. With a threshold of $\kappa=3$ this algorithm
clips 39 FM, 0 FO and 1 T2 Cepheid. For the FM Cepheids the parameters
of the fitted line and the dispersion are close to the those of the
MAD clipping method (c.f. Table \ref{table_PLRs}). This
is not surprising since the sample is the same but for one FM Cepheid
that is additionally clipped by the E14 method. Of
course the threshold was also chosen such that both methods perform
as identically as possible, while still using an integer value for the
threshold. If we would not require the threshold to be integer, we
could find a threshold that gives the same result as the MAD
clipping.  Using the same threshold for the FO Cepheids as for the FM
Cepheids results in no clipping at all. The MAD method
on the other hand does only require one threshold for all Cepheid
types. The convergence of the internal scatter method is very
sensitive to the threshold $\kappa$  and the starting value of $\sigma_{int}$ (we chose
$\sigma_{int}=0$). While the basic idea behind
both methods is the same, namely increasing the error by a constant
that is described by the dispersion, the method introduced by
E14 requires one additional free parameter and
according to our tests the convergence performance depends on the
starting parameters. The MAD clipping method on the
other hand does not depend on the starting parameters and is very easy to
implement.

\section{The adopted Period-Luminosity relations\label{chapter_PLRs}}

\begin{figure}
\epsscale{0.75}
\plotone{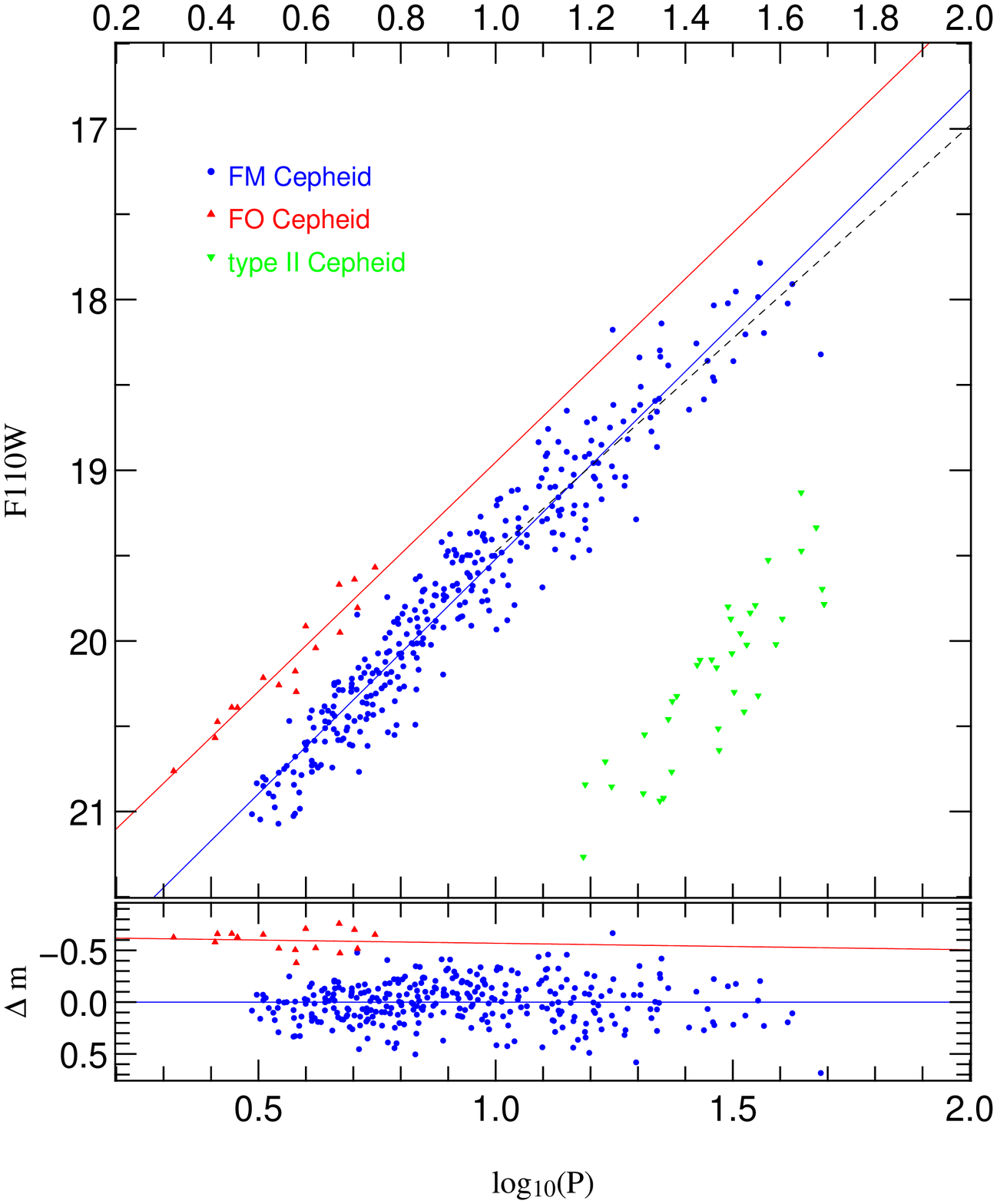}
\caption{F110W Period-Luminosity relation. The outlier rejection was
  performed with the MAD method in the Wesenheit PLR
  (Fig. \ref{fig_PLR_Wesenheit}). A Cepheid that was clipped in the
  Wesenheit PLR was rejected in the near-infrared bands. The fit
  parameters to the FM (blue solid line) and FO (red solid line) PLRs
  are given in Table \ref{table_PLRs} ($\#1$ for the FM and $\#3$ for
  the FO). The $\log(P)>1$ FM PLR is shown as a black dashed line
  ($\#2$ in Table \ref{table_PLRs}). The type II Cepheids show no
    linear relationship (see
    Section \ref{chapter_PLRs}) \label{fig_PLR_F110W}}
\end{figure}

\begin{figure}
\epsscale{0.75}
\plotone{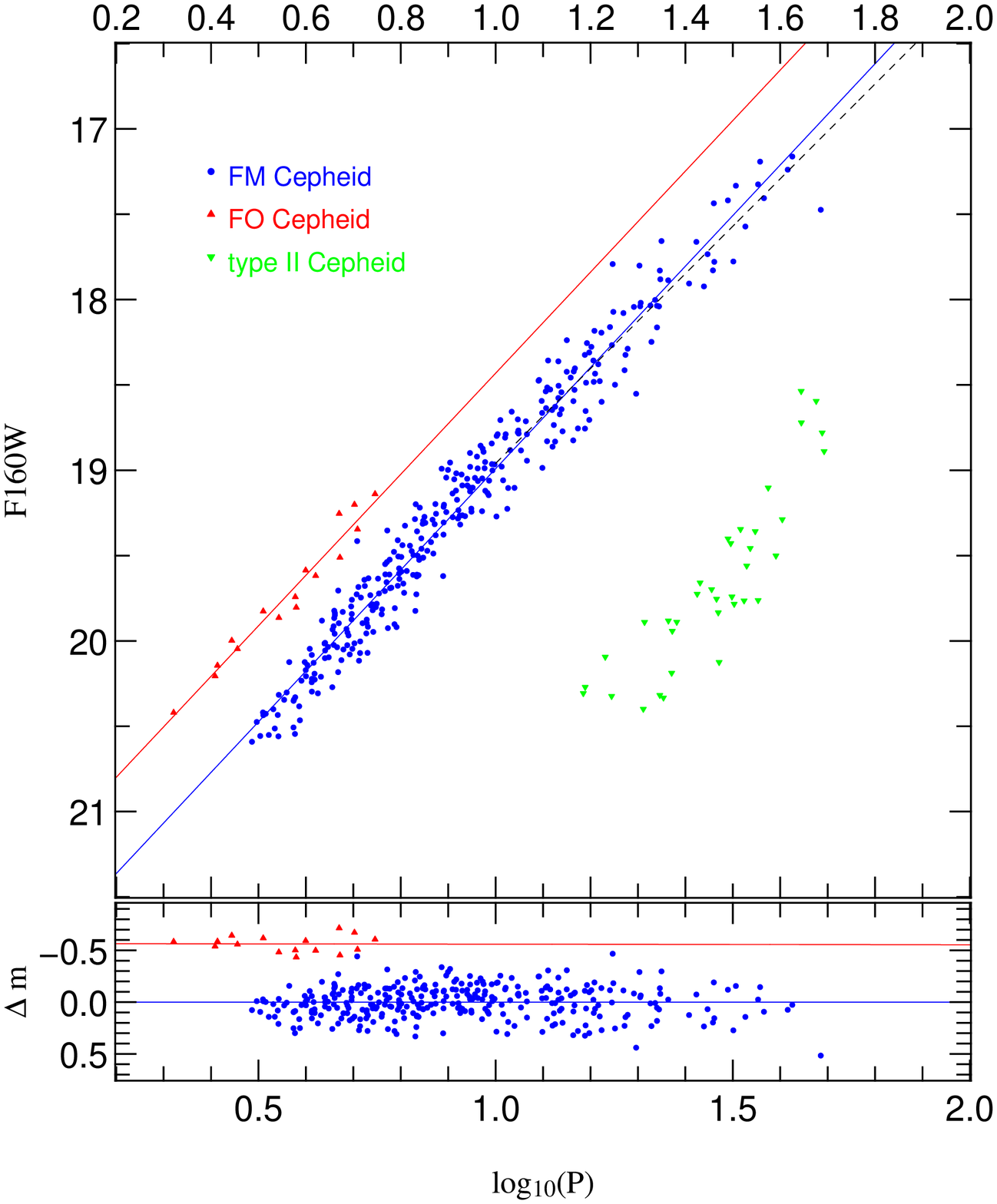}
\caption{F160W Period-Luminosity relation. The outlier rejection was
  performed with the MAD method in the Wesenheit PLR
  (Fig. \ref{fig_PLR_Wesenheit}). A Cepheid that was clipped in the
  Wesenheit PLR was rejected in the near-infrared bands. The fit
  parameters to the FM (blue solid line) and FO (red solid line) PLRs
  are given in Table \ref{table_PLRs} ($\#4$ for the FM and $\#6$ for
  the FO). The $\log(P)>1$ FM PLR is shown as a black dashed line
  ($\#5$ in Table \ref{table_PLRs}). The type II Cepheids show no
    linear relationship (see
    Section \ref{chapter_PLRs})}\label{fig_PLR_F160W}
\end{figure}

The F110W and F160W PLRs are shown in Fig. \ref{fig_PLR_F110W} and
Fig. \ref{fig_PLR_F160W}. Table \ref{table_PLRs} contains the
corresponding best fit parameters. The fits of the Wesenheit PLRs
shown in Fig. \ref{fig_PLR_Wesenheit} and
Fig. \ref{fig_Wesenheit_Efstathiou} are also included in this
table. The PLR fits are of the form $m = a + b \cdot \log(P)$ with a
dispersion of $\sigma$. $N_{fit}$ is the number of Cepheids
contributing to the fit and $\sigma_{int}$ is given for the cases
where the internal scatter clipping method was used \citep[see also
  Equation \ref{eqn_Efstathiou}]{Efstathiou}. We included the type II
Cepheids in the figures but do not fit a PLR since these do not appear
to show one clear linear relationship. Another reason not to fit
  a PLR is that the transition between W Vir stars and RV Tauri is at
  $\log(P)\approx 1.3$ and according to \citet{2009MNRAS.397..933M}
  the PLRs of both type II subgroups are not collinear. This can also
  be seen in the recent study of \citet{2015MNRAS.446.3034R} where the RV Tauri stars
  are not on the linear PLR of the other type II Cepheids.

The R12 PLRs with $b_{F110W}=-2.725~(0.150)$ and
$b_{F160W}=-3.003~(0.127)$ are steeper than our corresponding slopes
(the slopes for the $\log(P)>1$ subsample: $\#2$ and $\#5$ in Table
\ref{table_PLRs}). The Wesenheit slope cannot be compared since R12
use $R=1.54$ while we use a different value (c.f. Equations
\ref{eqn_Wesenheit_1} and \ref{eqn_Wesenheit_2}) derived from
\citet[table 6 with $R_V=3.1$]{2011ApJ...737..103S}. In fact the
slopes of the R12 sample are closer to our PLRs for the full
sample ($\#1$ and $\#4$ in Table \ref{table_PLRs}). Nevertheless the
slopes of both samples agree within their $1\sigma$ error bars.

The R12 PLR fits (Table 2 in R12) are slightly inconsistent to the PLR
Fig. 2 given in R12. Reanalyzing the R12 F160W data (with the double
entry of Cepheid vn.2.2.463 in the data as mentioned before) we can
reproduce the R12 slope but get an offset of 0.06 mag for
$m(\log(P)=1.2)$. This PLR is closer to the one shown in the R12 PLR
plot.

\begin{figure}
\epsscale{0.75}
\plotone{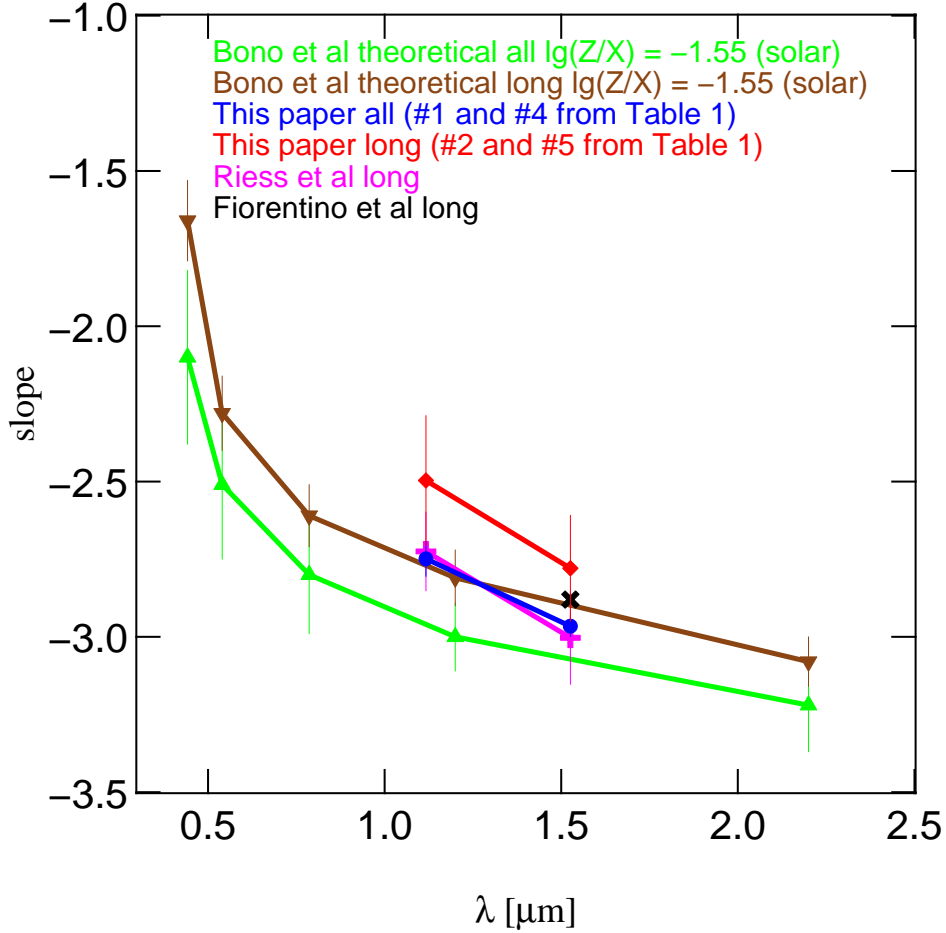}
\caption{PLR slope dependence on the wavelength. The R12
  slopes (magenta) are steeper than our slopes for the long period
  Cepheid sample ($\log(P)>1$, red, $\#2$ and $\#5$ in Table
  \ref{table_PLRs}). The \citet{2013MNRAS.434.2866F} slope with
  $Z=0.02$ and $Y=0.28$ for long period Cepheids (black) is
    within the error of the slope of  our long
  period sample result. For a different Helium content the
  \citet{2013MNRAS.434.2866F} slope is steeper and would agree with
  our total Cepheid sample (red, $\#1$ and $\#4$ in Table
  \ref{table_PLRs}). The theoretical predictions of
\citet{2010ApJ...715..277B} are steeper than our measurements for both
subsamples. \label{fig_compare_results}}
\end{figure}

The comparison of the slopes can also be
seen in Fig. \ref{fig_compare_results}. The theoretical predictions of
\citet{2010ApJ...715..277B} for the slopes of the different subsamples
are all steeper than our measurements or those of R12.
Unfortunately we cannot compare our results for the Wesenheit PLR with
E14 since they use a Wesenheit function that includes V and I
band magnitudes, which we do not have.

\begin{deluxetable}{cccccccccc}
\tabletypesize{\scriptsize}
\rotate
\tablecaption{PLR fit parameters\label{table_PLRs}}
\tablewidth{0pt}
\tablehead{
\colhead{$\#$} & \colhead{band} & \colhead{type} & \colhead{range} & \colhead{$N_{fit}$} &
\colhead{a (log P = 1)} & \colhead{slope b} & \colhead{$\sigma$} &
\colhead{$\sigma_{int}$\tablenotemark{a}} & \colhead{$\chi_{d.o.f.}^2$\tablenotemark{b}}
}
\startdata
 1 & F110W & FM & all & 319 & 19.521 ( 0.012) & -2.749 ( 0.057) &   0.204 & - &  1.000 \\
 2 & F110W & FM & log P $>$ 1 & 110 & 19.476 ( 0.037) & -2.497 ( 0.209) &   0.243 & - &  1.415 \\
 3 & F110W & FO & all & 16 & 18.953 ( 0.051) & -2.686 ( 0.157) &   0.105 & - &  1.000 \\
 4 & F160W & FM & all & 319 & 18.991 ( 0.003) & -2.966 ( 0.033) &   0.155 & - &  1.000 \\
 5 & F160W & FM & log P $>$ 1 & 110 & 18.960 ( 0.028) & -2.779 ( 0.171) &   0.178 & - &  1.318 \\
 6 & F160W & FO & all & 16 & 18.431 ( 0.051) & -2.960 ( 0.145) &   0.082 & - &  1.000 \\
 7 & Wesenheit & FM & all & 319 & 18.255 ( 0.007) & -3.267 ( 0.071) &   0.138 & - &  1.000 \\
 8 & Wesenheit & FM & log P $>$ 1 & 110 & 18.244 ( 0.016) & -3.172 ( 0.117) &   0.147 & - &  1.145 \\
 9 & Wesenheit & FO & all & 16 & 17.708 ( 0.134) & -3.339 ( 0.281) &   0.074 & - &  1.000 \\
 10 & Wesenheit & FM & all & 318 & 18.256 ( 0.004) & -3.270 ( 0.036) &   0.136 &  0.128 &  1.126 \\
 11 & Wesenheit & FM & log P $>$ 1 & 110 & 18.244 ( 0.016) & -3.172 ( 0.117) &   0.147 &  0.128 &  1.183 \\
 12 & Wesenheit & FO & all & 19 & 17.705 ( 0.135) & -3.414 ( 0.282) &   0.265 &  0.265 &  1.062 \\
\enddata
\tablenotetext{a}{internal scatter as defined by E14}
\tablenotetext{b}{reduced $\chi^2$}
\tablecomments{The magnitude errors were set to the same value, namely to the
  dispersion $\sigma$. In the cases where the E14 clipping
  method was used ($\#10$, $\#11$ and $\#12$) the internal scatter
  ($\sigma_{int}$) was added in quadrature to the photometric errors
  (c.f. Fig. \ref{fig_Wesenheit_Efstathiou}). The errors of the fitted
  parameters were determined with the bootstrapping method. Lines
  8 and 11 show identical parameters since the only
  difference is that the magnitude errors for line 11 include the
  photometric errors determined by DOLPHOT which as mentioned earlier
  are negligible compared to $\sigma_{int}$.
}
\end{deluxetable}

\begin{figure}
\epsscale{0.75}
\plotone{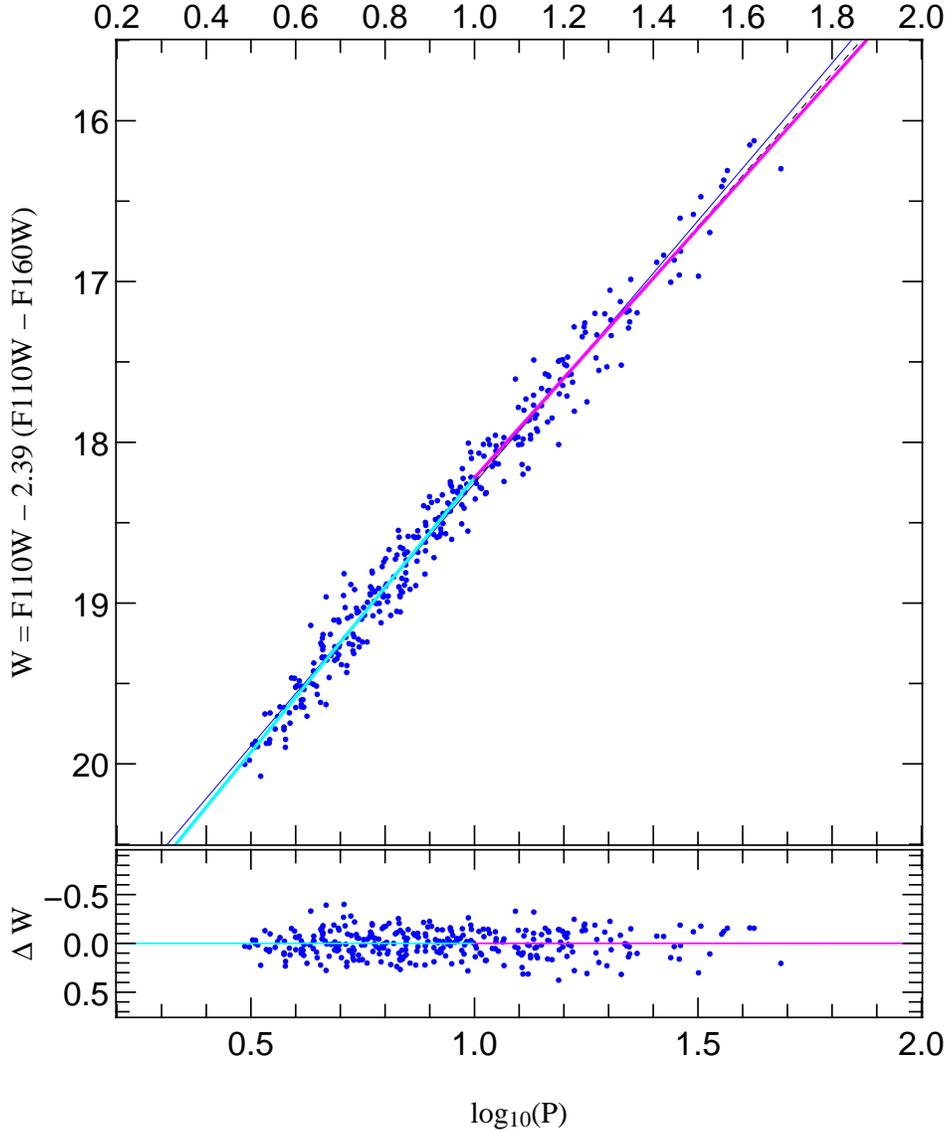}
\caption{Broken slope Wesenheit PLR for FM Cepheids. A broken slope
  fit with a common suspension point at 10 days (c.f. Equations
  16 and 17 in K13) is shown ($\#1$ in Table \ref{table_PLRs_broken}). The short period Cepheid slope
  ($\log(P)\leq1$) is shown in cyan and the long period Cepheid slope
  ($\log(P)>1$) in magenta. The blue solid line is the linear slope
  fit ($\#7$ in Table \ref{table_PLRs}) and the black dashed line the
  fit to the long period Cepheid sample ($\#8$ in Table \ref{table_PLRs}).  \label{fig_PLR_Wesenheit-suspension}}
\end{figure}

\begin{figure}
\epsscale{1.0}
\plotone{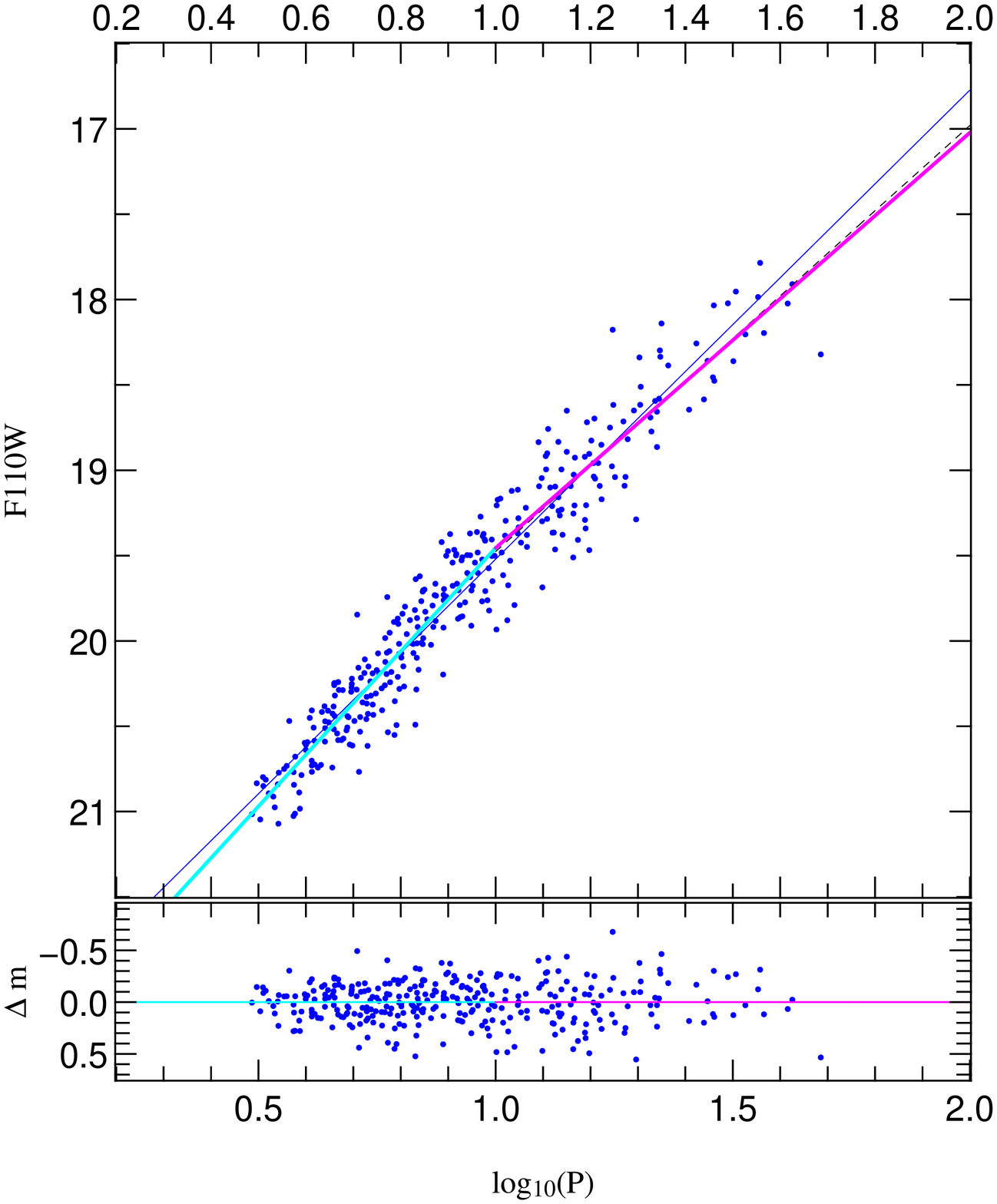}
\caption{Broken slope F110W PLR for FM Cepheids ($\#2$ in Table
  \ref{table_PLRs_broken}). Same as
  Fig. \ref{fig_PLR_Wesenheit-suspension} but with $\#1$ (Table
  \ref{table_PLRs}) as linear slope fit and $\#2$ (Table
  \ref{table_PLRs}) as fit for the long period
  Cepheids. \label{fig_PLR_F110W-suspension}}
\end{figure}

\begin{figure}
\epsscale{1.0}
\plotone{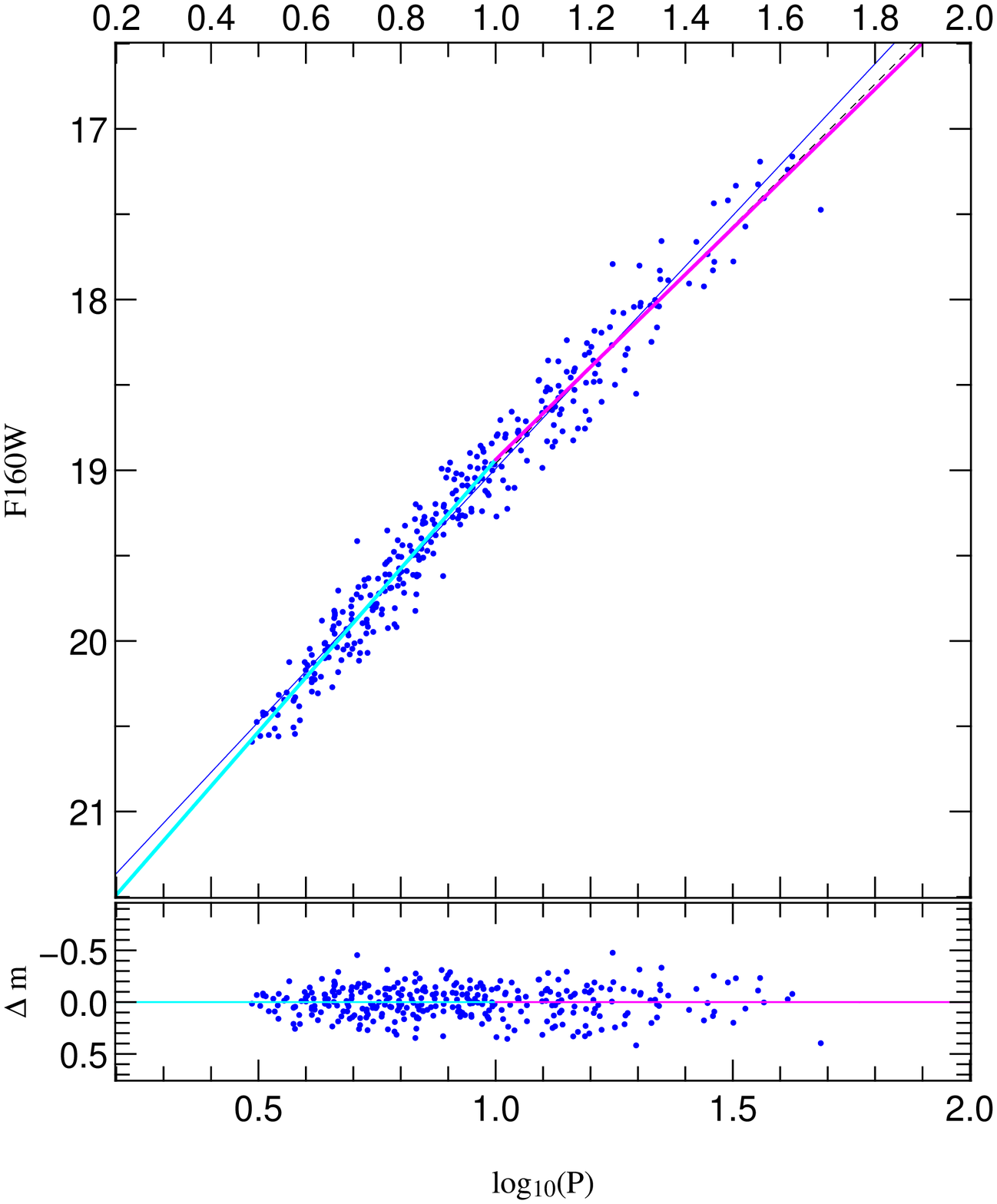}
\caption{Broken slope F160W PLR for FM Cepheids ($\#3$ in Table
  \ref{table_PLRs_broken}). Same as
  Fig. \ref{fig_PLR_Wesenheit-suspension} but with $\#4$ (Table
  \ref{table_PLRs}) as linear slope fit and $\#5$ (Table
  \ref{table_PLRs}) as fit for the long period
  Cepheids. \label{fig_PLR_F160W-suspension}}
\end{figure}

\begin{figure}
\epsscale{1.0}
\plotone{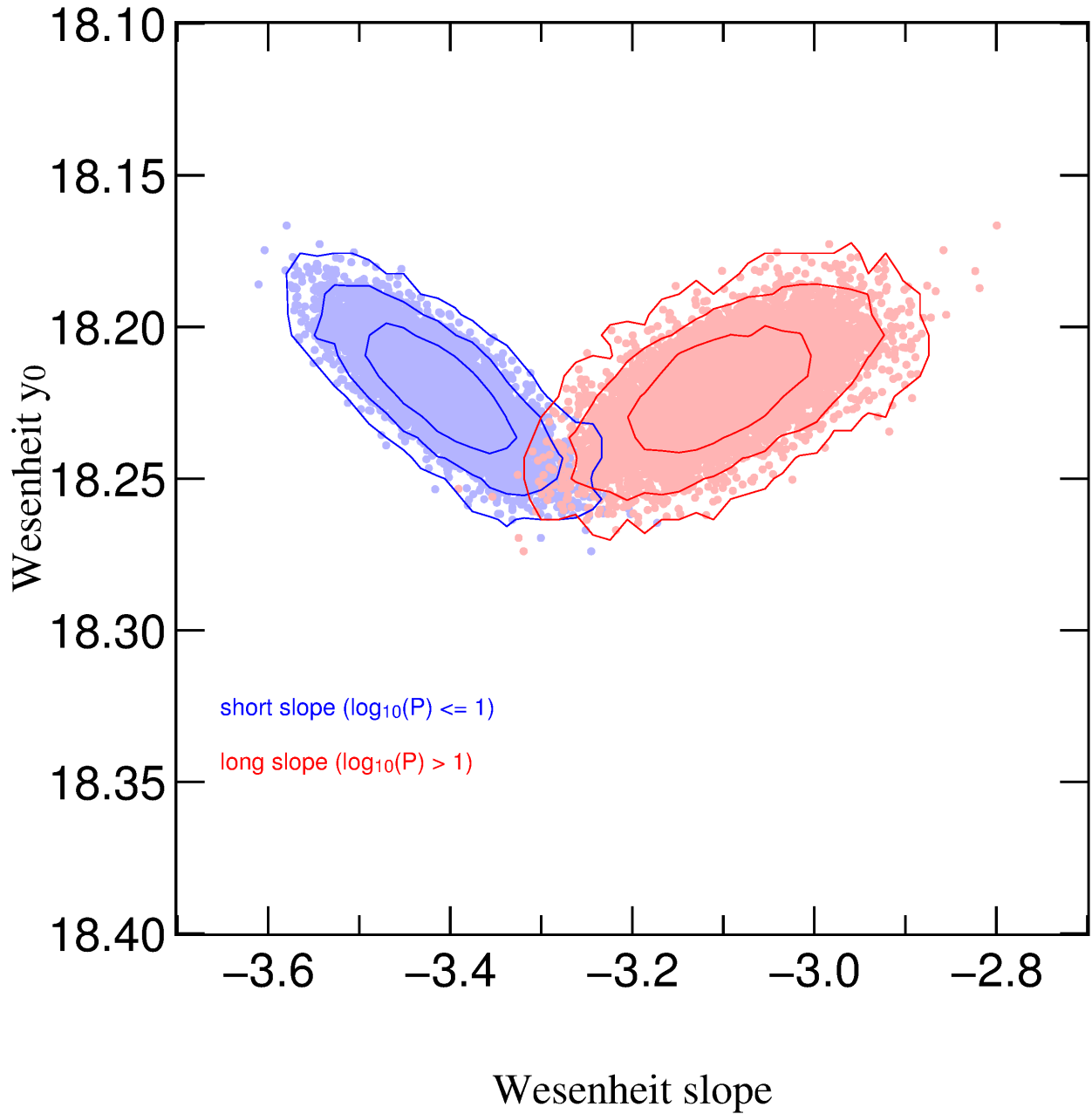}
\caption{Bootstrapping of the broken slope in the Wesenheit PLR. The
  common suspension point $y_0$ is plotted versus the slope. The short
  period Cepheid slope ($\log(P)\leq1$) is shown in blue and the long
  period Cepheid slope in red. The 1$\sigma$, 2$\sigma$ and 3$\sigma$
  contour lines are also shown as solid
  lines.\label{fig_PLR_Wesenheit-suspension_bootstraping}}
\end{figure}

\begin{figure}
\epsscale{1.0}
\plotone{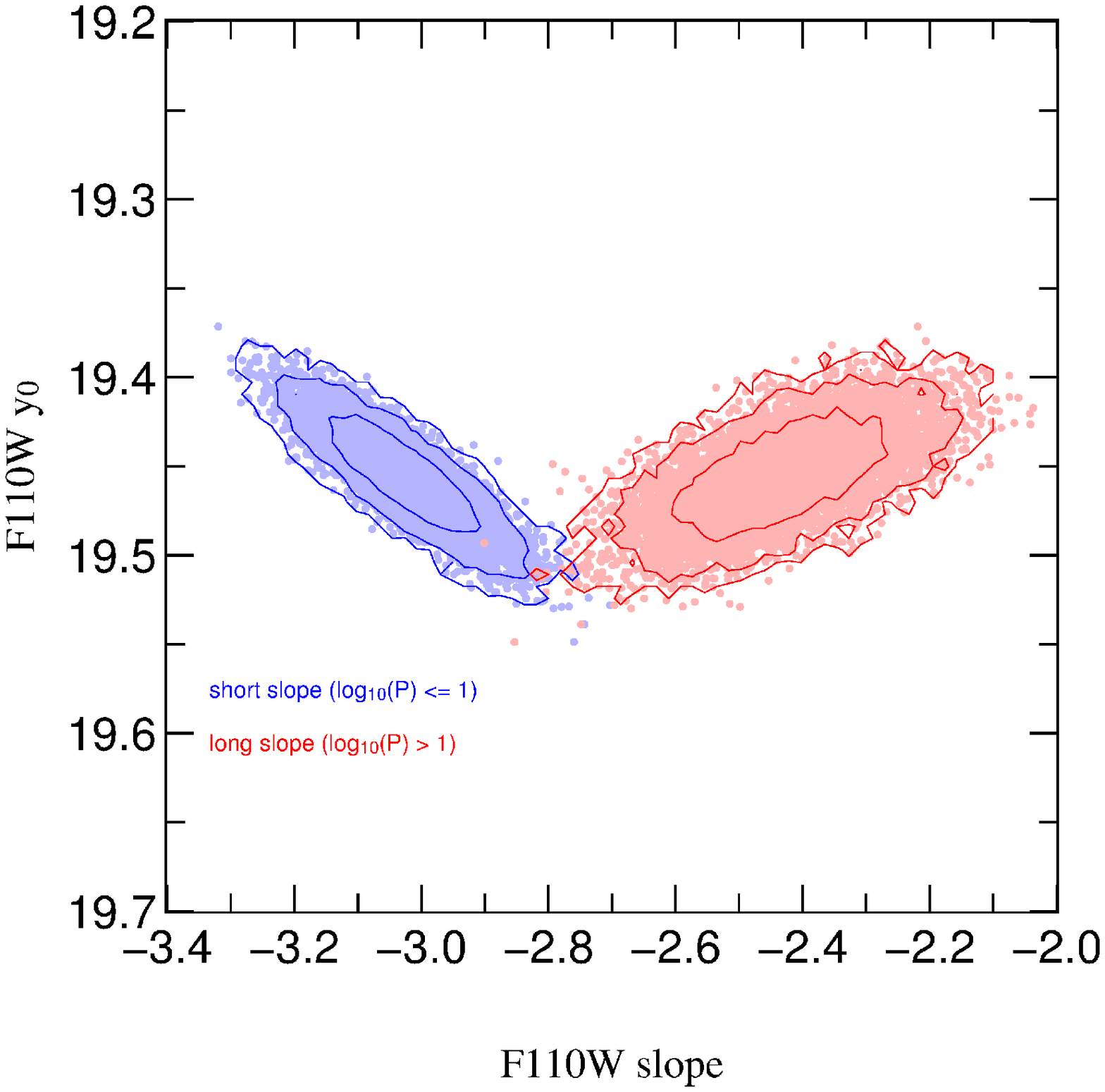}
\caption{Bootstrapping of the broken slope in the F110W PLR. The legend is
  otherwise similar to that displayed in
  Fig. \ref{fig_PLR_Wesenheit-suspension_bootstraping}.\label{fig_PLR_F110W-suspension_bootstraping}}
\end{figure}

\begin{figure}
\epsscale{1.0}
\plotone{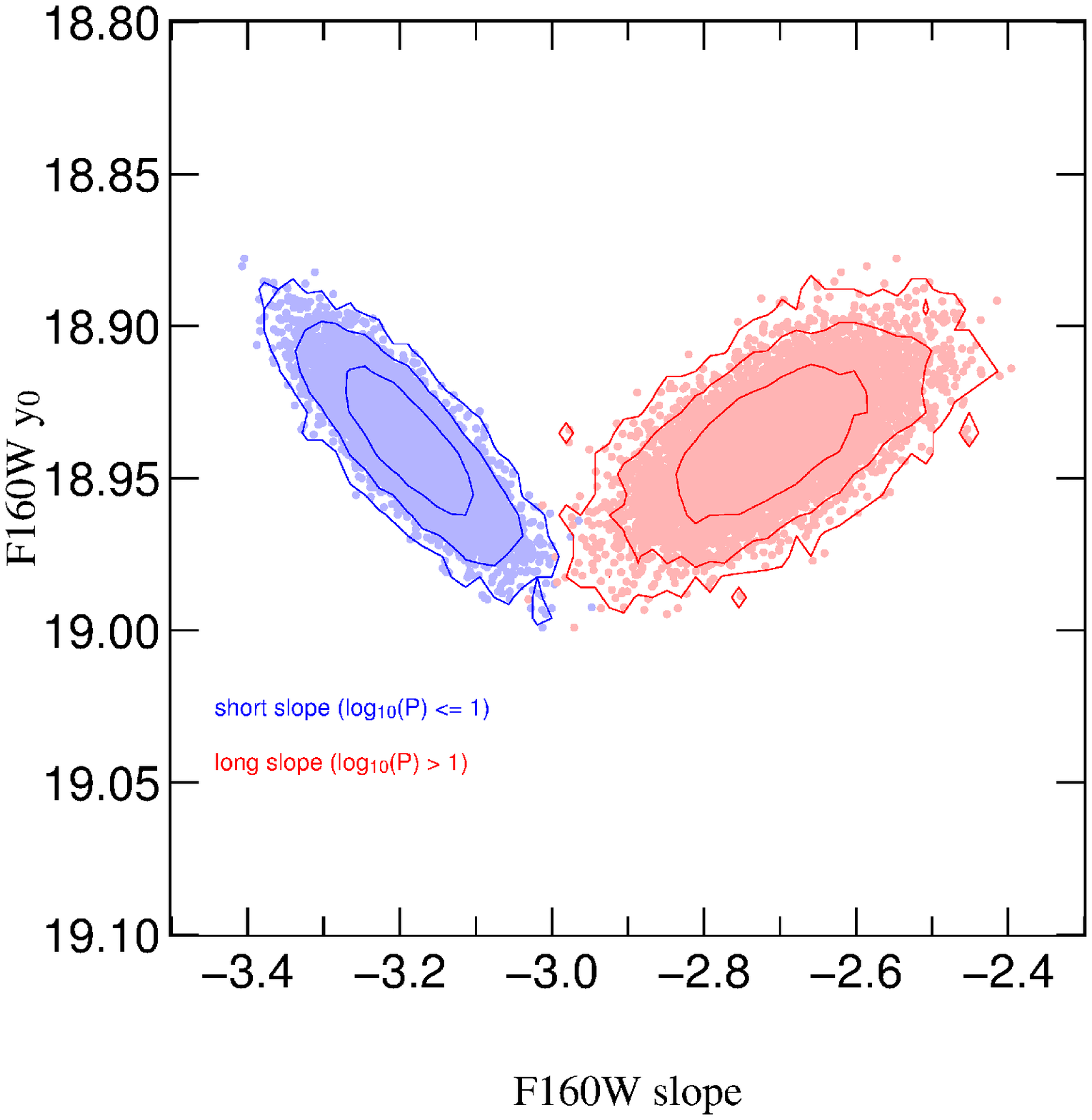}
\caption{Bootstrapping of the broken slope in the F160W PLR. The legend is
  otherwise similar to that displayed in
  Fig. \ref{fig_PLR_Wesenheit-suspension_bootstraping}.\label{fig_PLR_F160W-suspension_bootstraping}}
\end{figure}

In the next step we investigate whether our FM Cepheids show any
signature of the broken slope proposed by \citet{Sandage}. For this we
use the same approach as in Equations 16 and 17 in K13: We fit two
slopes and a common suspension point at 10 days. These fits can be
seen in Fig. \ref{fig_PLR_Wesenheit-suspension},
Fig. \ref{fig_PLR_F110W-suspension} and
Fig. \ref{fig_PLR_F160W-suspension}. The fit parameters are summarized
in Table \ref{table_PLRs_broken}. All fits show a steeper slope for
short period Cepheids ($\log(P)\leq1$) than for long period Cepheids
($\log(P)>1$) .  Note that a Malmquist bias would influence the faint
end slope so that it becomes shallower than it actually is. We also
perform bootstrapping (resample the data) with 10000
realizations to check how significant the broken slope is and show the
results in Fig. \ref{fig_PLR_Wesenheit-suspension_bootstraping},
\ref{fig_PLR_F110W-suspension_bootstraping} and
\ref{fig_PLR_F160W-suspension_bootstraping}. Only for the Wesenheit
function there are realizations of the bootstrapping where the 3
$\sigma$ contours overlap. The break at exactly 10 days is often
  adopted in the literature, but there are also studies contesting
  that value. \citet{2009A&A...504..959K} for example find that the
  break occurs at 10.47 days. We discuss the break at 10 days here and
  provide a table for the relevant parameters of other suspension
  points in the appendix.

\begin{deluxetable}{ccccccc}
  \tabletypesize{\scriptsize} \rotate \tablecaption{broken slope PLR
    fit parameters\label{table_PLRs_broken}} \tablewidth{0pt} \tablehead{
    \colhead{$\#$} & \colhead{band} & \colhead{$b_{\log(P)\leq1}$} &
    \colhead{$b_{\log(P)>1}$} &\colhead{$a_{\log(P)=1}$} &
    \colhead{$\sigma$} & \colhead{$\chi_{d.o.f.}^2$\tablenotemark{a}}
}
\startdata
 1 & Wesenheit & -3.411 ( 0.038) & -3.103 ( 0.060) & 18.221 ( 0.013) &  0.136 &  0.978 \\
 2 & F110W & -3.028 ( 0.078) & -2.433 ( 0.105) & 19.455 ( 0.021) &  0.200 &  0.960 \\
 3 & F160W & -3.188 ( 0.050) & -2.714 ( 0.069) & 18.938 ( 0.014) &  0.152 &  0.956 \\
\enddata
\tablenotetext{a}{reduced $\chi^2$}
\tablecomments{The magnitude errors were set to the same value, namely to the
  dispersion $\sigma$ of the corresponding fit in Table
  \ref{table_PLRs} ($\#7$, $\#1$ and $\#4$). The errors of the parameters were determined with bootstrapping.}
\end{deluxetable}

The results from the bootstrapping already point toward a broken
slope. To determine if the broken slope is significantly better than
the linear slope we perform an F-test. The advantage of the F-test is
that it is not sensitive to the problem of the uncertainty in the
adopted magnitude errors. Due to the fact that we chose the magnitude
errors to be equal to the dispersion in the linear fit, we are able to
get better estimates on the errors of the fitted parameters. However, this
approach does not allow us to perform a $\chi^2$ test. Following
Equations 3.40 and 3.41 from \citet{chatterjee2013regression} where
model 1 is the reduced model with $p_1$ parameters and model 2
the full model with $p_2$ parameters, the observed
F-ratio is:
\begin{equation}
F_{obs}=\frac{[\chi_1^2 - \chi_2^2]/[p_2-p_1]}{\chi_2^2/[N-p_2]}
\label{eqn_Ftest}
\end{equation}
$N$ denotes the number of data points and the $\chi_i^2$ are the
corresponding $\chi^2$ of the two models. The critical F-value is
\begin{equation}
F_{crit}=F(p_2-p_1,N-p_2;\alpha)
\label{eqn_Ftest_crit}
\end{equation}
for a significance level of $\alpha$, where F is the distribution function
of the F-test. For $F_{obs} \geq F_{crit}$ the
null hypothesis (that model 2 is not significantly better than model
1) is rejected. Simply put for $F_{obs} \geq F_{crit}$ model 2 is more
significant than model 1. In our case model 1 is the linear fit (Table
\ref{table_PLRs}) and model 2 the fit with the broken slope (Table
\ref{table_PLRs_broken}). For a typical significance level of
$\alpha=0.05$ the critical F-value is 
$F_{crit}=F(1,316;0.05)=3.87$. Our observed F-values are 
$F_{obs,Wesenheit}=8.24$, $F_{obs,F110W}=14.12$ and
$F_{obs,F160W}=15.71$. Therefore all three broken slope fits are significant
at a level of at least $1-\alpha=0.95$. We confirm the result from the
bootstrapping i.e. the Wesenheit broken slope is less significant than
the F110W and F160W broken slopes. Indeed the F110W and F160W broken
slopes are also still significant at a 3 $\sigma$ level.

\begin{figure}
\epsscale{1.0}
\plotone{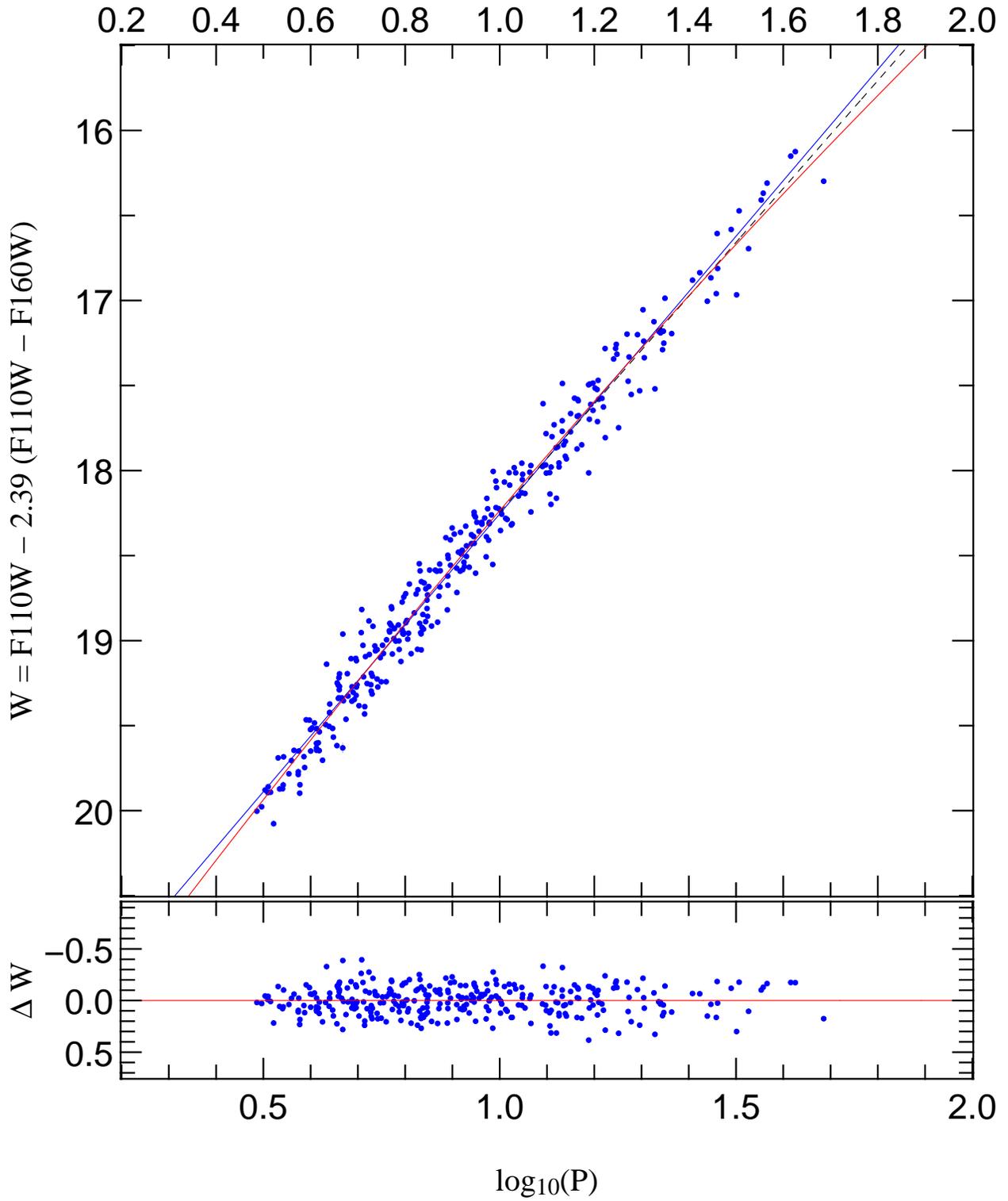}
\caption{Parabolic Wesenheit PLR for FM Cepheids ($\#1$ in Table
  \ref{table_PLRs_parabola}). The parabola is shown as a red solid
  line. The blue solid line is the linear slope fit ($\#7$ in Table
  \ref{table_PLRs}) and the black dashed line the
  fit to the long period Cepheid sample ($\#8$ in Table
  \ref{table_PLRs}). \label{fig_PLR_Wesenheit-parabola}}
\end{figure}

\begin{figure}
\epsscale{1.0}
\plotone{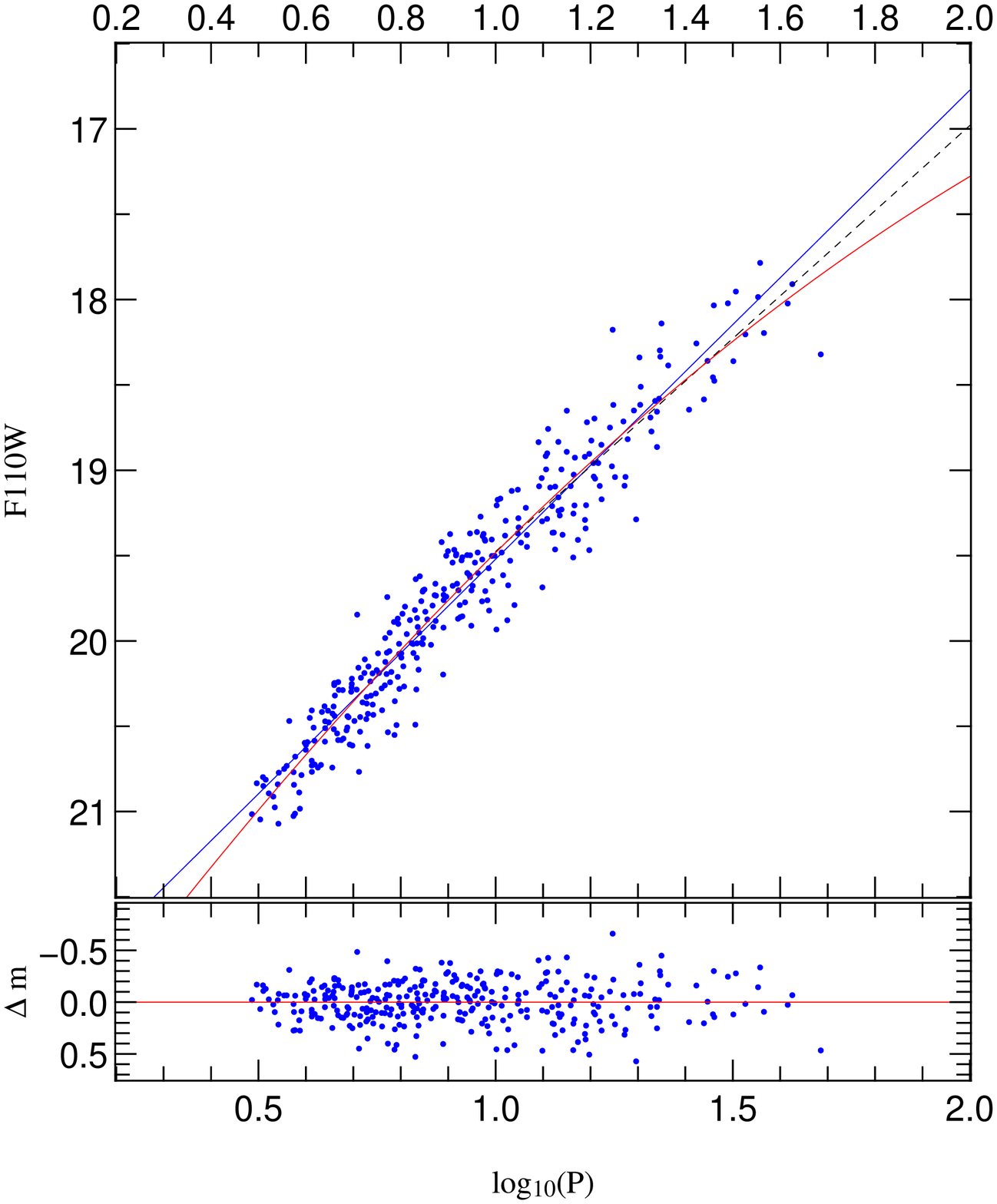}
\caption{Parabolic F110W PLR for FM Cepheids (red solid line, $\#2$ in
  Table \ref{table_PLRs_parabola}). The blue solid line is the linear
  slope fit ($\#1$ in Table \ref{table_PLRs}) and the black dashed
  line the fit to the long period Cepheid sample ($\#2$ in Table
  \ref{table_PLRs}).\label{fig_PLR_F110W-parabola}}
\end{figure}

\begin{figure}
\epsscale{1.0}
\plotone{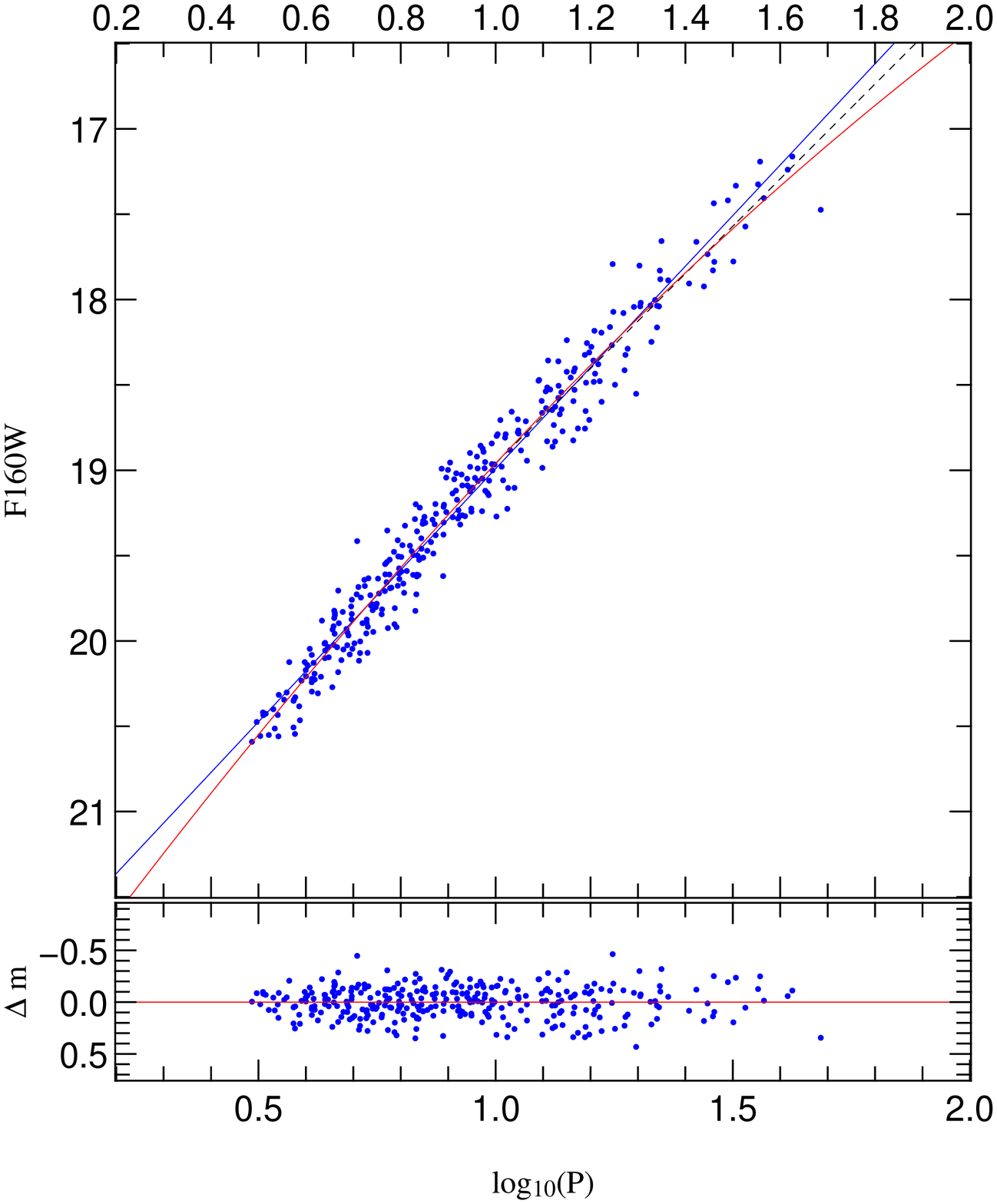}
\caption{Parabolic F160W PLR for FM Cepheids (red solid line, $\#3$ in
  Table \ref{table_PLRs_parabola}). The blue solid line is the linear
  slope fit ($\#4$ in Table \ref{table_PLRs}) and the black dashed
  line the fit to the long period Cepheid sample ($\#5$ in Table
  \ref{table_PLRs}). \label{fig_PLR_F160W-parabola}}
\end{figure}

In the next step we check how well the data are described by a
parabola instead of a broken linear relation. The parabolic fits
  are shown in Fig. \ref{fig_PLR_Wesenheit-parabola},
  \ref{fig_PLR_F110W-parabola} and \ref{fig_PLR_F160W-parabola}. The
  fit parameters are summarized in Table \ref{table_PLRs_parabola}. As
  can already be seen from the $\sigma_{d.o.f.}$ the parabolic fit
  will practically be as significant as the broken slope.

\begin{deluxetable}{ccccccc}
  \tabletypesize{\scriptsize} \rotate \tablecaption{parabola PLR
    fit parameters\label{table_PLRs_parabola}} \tablewidth{0pt} \tablehead{
    \colhead{$\#$} & \colhead{band} & \colhead{$a_{\log(P)=1}$} &
    \colhead{$b$} &\colhead{$c$} &
    \colhead{$\sigma$} & \colhead{$\chi_{d.o.f.}^2$\tablenotemark{a}}
}
\startdata
 1 & Wesenheit & 18.236 ( 0.011) & -3.265 ( 0.029) &  0.267 ( 0.089) &  0.136 &  0.980 \\
 2 & F110W & 19.482 ( 0.016) & -2.746 ( 0.039) &  0.543 ( 0.141) &  0.200 &  0.960 \\
 3 & F160W & 18.960 ( 0.012) & -2.964 ( 0.032) &  0.427 ( 0.107) &  0.152 &  0.957 \\
\enddata
\tablenotetext{a}{reduced $\chi^2$}
\tablecomments{The magnitude errors were set to the same value, namely to the
  dispersion $\sigma$ of the corresponding fit in Table
  \ref{table_PLRs} ($\#7$, $\#1$ and $\#4$) and the errors of the fit parameters determined with bootstraping.
  The parabola fit has the form $m=a + b \cdot \log(P) + c \cdot [\log(P)]^2$.}
\end{deluxetable}

A possible reason for the broken slope could be the Hertzsprung
progression. With increasing periods the bump in the light curve moves
to the maximum (brightest magnitude) of the light curve. For periods
larger than 10 days the bump moves away from the maximum (see
e.g. K13). Randomly phased observations might be biased toward
brighter magnitudes due to the bump in the light curve. This effect
would be strongest for Cepheids around 10 days and for larger periods
it would decrease again. This would mean that the magnitudes at 10
days are systematically brighter than they should be, which could
explain the broken slope.

In the light curves published in \citet{2004AJ....128.2239P} we see
that the bumps are stronger in the J band than in the H band. This
fits to our result that in the F110W band (close to the J band) the
curvature of the parabolic fit to the PLR is stronger than in the
F160W band (close to H band). Also the decrease of the slope of the
long-period Cepheid PLR as compared to the linear fit to the full
sample is stronger in F110W than in F160W. In view of this, an
overestimation of the mean magnitudes of Cepheids near $\log(P)=1$
from random-phase data due to the bump presence seems indeed to be a
plausible explanation for the observed non-linearity in the PLRs or at
least is contributing to this effect. With the full PAndromeda data
set of three years we will be able to perform a phase correction and therefore be
capable to test whether such a hypothetical bias exists.

\clearpage

\section{Implications of the improved PLR on the Hubble constant\label{chapter_H0}}
 
As can be seen in Fig. \ref{fig_compare_results} and e.g.  in $\#5$ in
Table \ref{table_PLRs} our PLR is different from the R12 PLR. Therefore we
want to explore what impact our new sample has on the estimate of the Hubble constant
$H_0$.

We use method $\#10$ table 3 in R12 where M31 is used as the anchor for the
comparison. If we were to use a different fit where M31 only
contributes to the fit of the slope, we would have to do the complete
analysis of the SN Ia data. So the idea is to only check for relative
changes in the anchor and assume nothing changes in the SN Ia galaxy
analysis, i.e. plug our sample in as an anchor and leave everything
else the same. Furthermore we have to make the reasonable assumption
that the photometric offsets between our sample and the R12 sample are well
understood and described by:
\begin{eqnarray}
< \Delta m_{\mathrm{F160W}} > = < m_{\mathrm{F160W,R12}} - m_{\mathrm{F160W,K14}} > = -0.019~ \mathrm{mag} \label{eqn_diff_160}\\
< \Delta m_{\mathrm{F110W}} > = < m_{\mathrm{F110W,R12}} - m_{\mathrm{F110W,K14}} > = -0.258~ \mathrm{mag} \label{eqn_diff_110}
\end{eqnarray}
We have to make this assumption so that we can later compare the
offsets between the two samples.

\begin{figure}
\epsscale{1.0}
\plotone{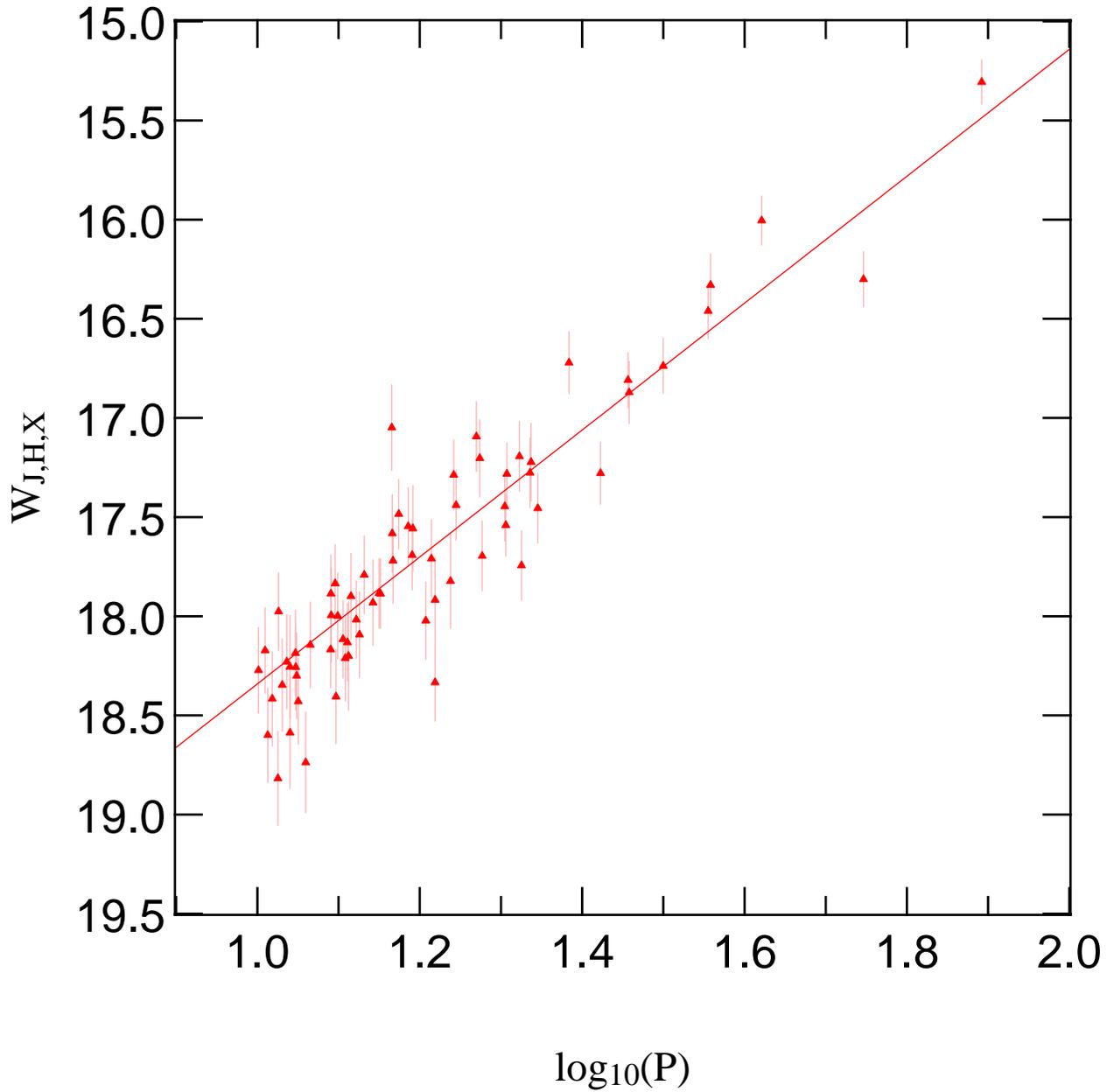}
\caption{Color corrected Wesenheit ($W_{\mathrm{J,H,X}}=\mathrm{F160W}-1.54(\mathrm{F110W}-\mathrm{F160W}+0.066)$) of the R12
  sample. The solid red line shows a fit of the slope of -3.20 ($\#
  10$ in table 3 R12) to the data. The fit gives a $m(\log(P)=1.2)_{\mathrm{R12}}=17.701~\mathrm{mag}$. \label{fig_method0_3}}
\end{figure}

\begin{figure}
\epsscale{1.0}
\plotone{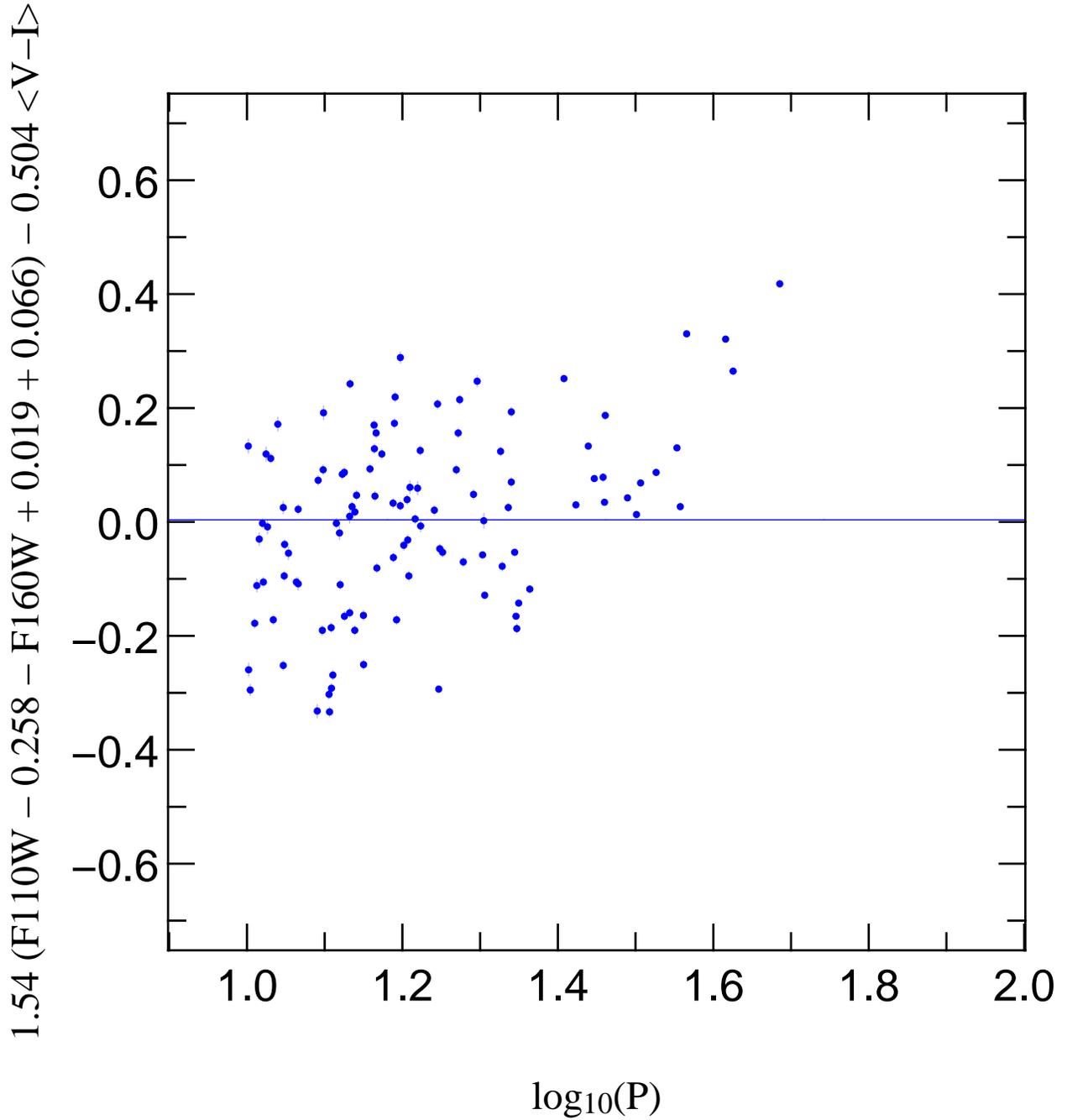}
\caption{The plot shows how well the color correction factor of
  $X=-0.066~\mathrm{mag}$ from R12 fits to our data that was
  transformed to the same photometric system used in R12 (i.e. with
  applied offsets). The blue solid line shows the mean offset of $0.004~\mathrm{mag}$.
  The brightness trend is also present in R12.  \label{fig_method1_2}}
\end{figure}

\begin{figure}
\epsscale{1.0}
\plotone {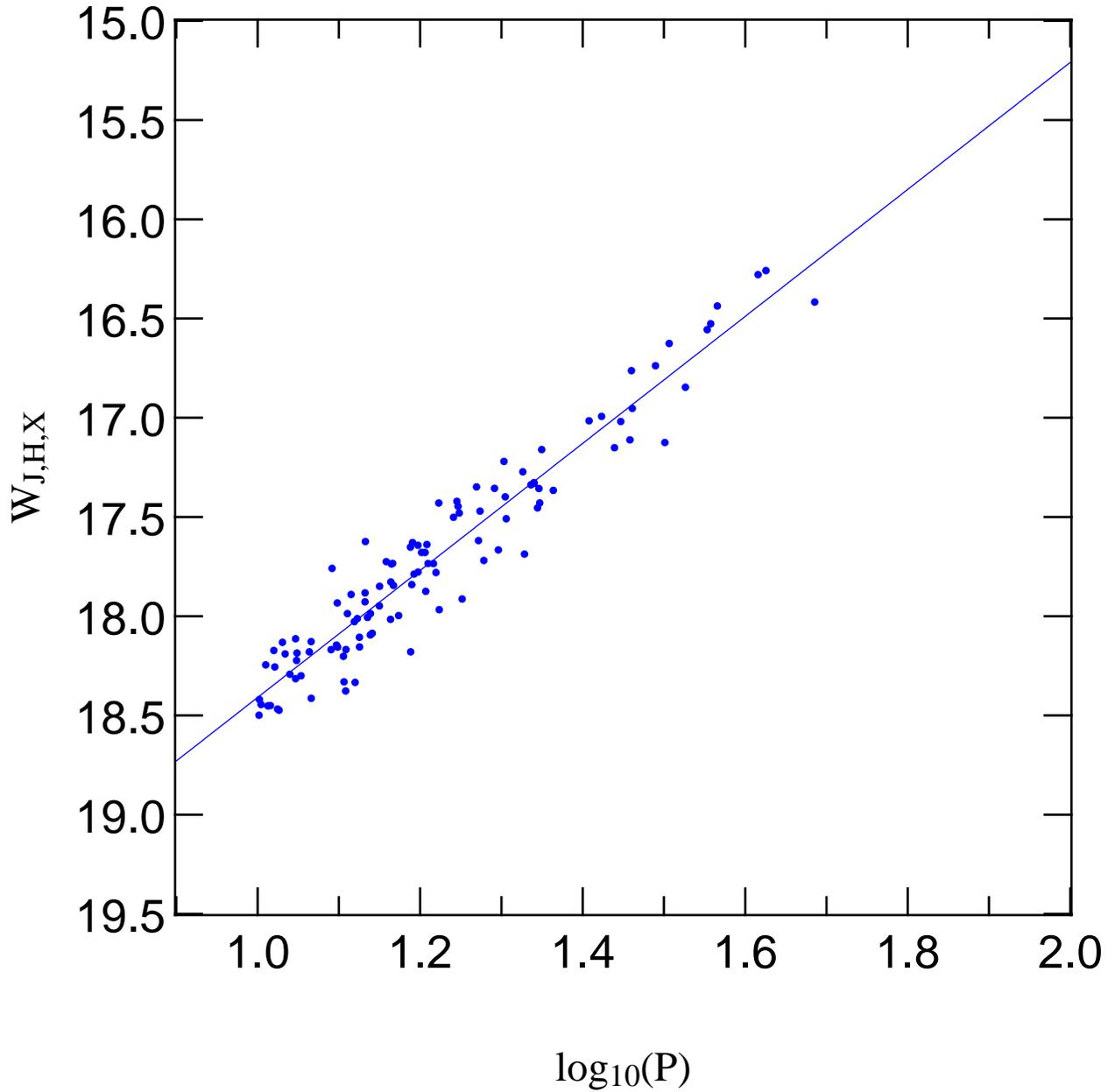}
\caption{Color corrected Wesenheit ($W_{\mathrm{J,H,X}}=\mathrm{F160W}-0.019-1.54(\mathrm{F110W}-0.258-\mathrm{F160W}+0.019+0.066)$) for our
  sample. Same as in Fig. \ref{fig_method0_3} the solid line shows a
  fit of the -3.20 slope to the data. The fit gives $m(\log(P)=1.2)
  _{\mathrm{K14}}=17.769~\mathrm{mag}$, which means that our sample
  is fainter than the R12 sample. \label{fig_method1_3}}
\end{figure}

The first step is to fit the color corrected Wesenheit function of the R12
sample with a slope of -3.20 as given by $\#10$ Table 3 in R12 in order
to obtain $m(\log(P)=1.2) _{\mathrm{R12}}$. Fig. \ref{fig_method0_3} shows the
fit to the color corrected Wesenheit function. In the next step we check how
well the color correction factor of $X=-0.066~\mathrm{mag}$ in R12
applies to our data. As can be seen in Fig. \ref{fig_method1_2} the color
correction factor is also consistent with our sample (the mean offset
is only $0.004~\mathrm{mag}$) when we apply the offsets described in
Equations \ref{eqn_diff_160} and \ref{eqn_diff_110}. The last step
is to fit the color corrected Wesenheit function with the offsets in order to obtain
$m(\log(P)=1.2) _{\mathrm{K14}}$. The fit shown in Fig. \ref{fig_method1_3} was done with
the same slope that was used in the first step. Due to the small
photometric errors in our sample the individual data points were not
weighted by their errors.

The magnitude difference for the two anchor samples is
\begin{equation}
\Delta M  = < m(\log(P)=1.2)_{\mathrm{R12}} - m(\log(P)=1.2)_{\mathrm{K14}} > =
17.701~\mathrm{mag} - 17.769~\mathrm{mag} = -0.068~\mathrm{mag}
\label{h0eqn3}
\end{equation}

This corresponds to
\begin{equation}
\Delta \mu = (m-M)_{\mathrm{R12}} - (m-M)_{\mathrm{K14}} = \Delta m - \Delta M = -
\Delta M = 0.068~\mathrm{mag}
\label{h0eqn4}
\end{equation}
i.e. only the difference in the anchor is relevant, since $ \Delta m =
0~\mathrm{mag} $ due to the first assumption.
Since $d_L \sim 1/H_0$ we get
\begin{equation}
\Delta \mu_0 = \mu_{\mathrm{0,R12}} - \mu_{\mathrm{0,K14}} = 5 \log\left(\frac{d_{\mathrm{L,R12}}}{d_{\mathrm{L,K14}}}\right) = 5 \log\left(\frac{H_{\mathrm{0,K14}}}{H_{\mathrm{0,R12}}}\right)
\label{eqn_H0_1}
\end{equation}
and therefore 
\begin{equation}
H_{\mathrm{0,K14}} = 10^{[\Delta \mu_0 / 5]} \cdot H_{\mathrm{0,R12}} = 1.032 \cdot H_{\mathrm{0,R12}}
\label{eqn_H0_2}
\end{equation}
 
\begin{figure}
\epsscale{1.0}
\plotone {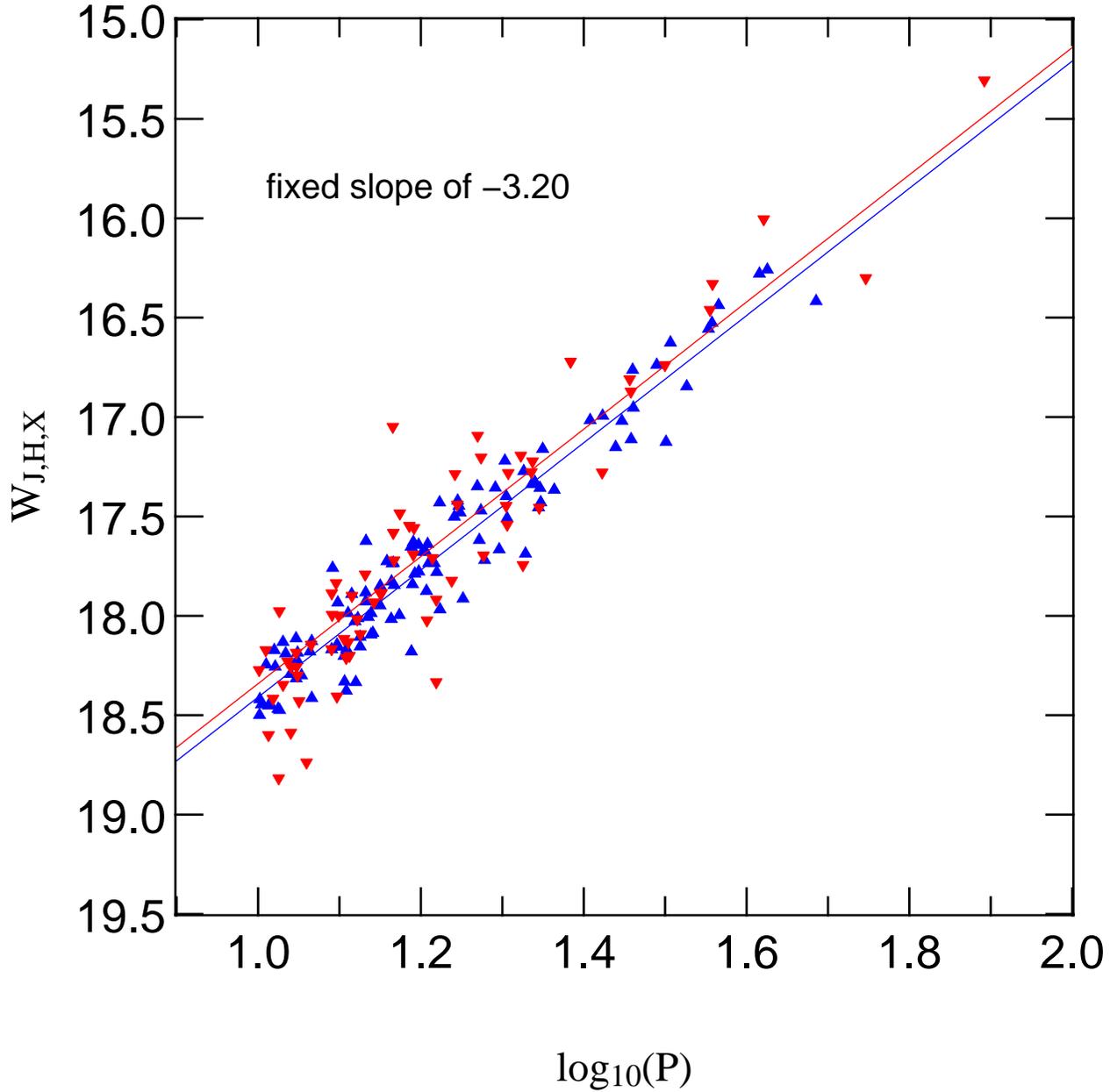}
\caption{Comparison between Fig. \ref{fig_method0_3} and
  Fig. \ref{fig_method1_3}. The blue triangles are our
  Cepheids, while the inverted red triangles are the R12
  Cepheids. The Wesenheit $W_{\mathrm{J,H,X}}$ is the color corrected
  Wesenheit function as defined in Fig. \ref{fig_method0_3} and
  Fig. \ref{fig_method1_3} for the respective samples (i.e. the blue
  Cepheids have the offsets applied as defined in Equations
  \ref{eqn_diff_160} and \ref{eqn_diff_110}). Both fits use a fixed
  slope of -3.20 that was also used in R12 for the SN Ia host
  galaxies.  \label{fig_final_H0_compare}}
\end{figure}

So our sample gives a $3.2\%$ increased $H_0$ compared to the R12
sample. This is very surprising when considering that the R12 sample
is in large part a subset of our
sample. Fig. \ref{fig_final_H0_compare} shows the difference in the
two samples.

 We checked if there is any indication
for this difference in the spatial distribution of the two data sets,
since our sample covers more of the M31 area.  But the subsets are
distributed equivalently across M31. It is not the case that one
subsample is located in the spiral arms and the other is not. As can
be seen in the appendix the crowding tests also support the argument
that the spatial distribution is not the reason for the offset, since
the offset only changes slightly.

The offset that is described in this section is very worrisome since
it begs the question how well we can constrain $H_0$ if a larger
Cepheid sample that covers more of the galaxy produces a different
$H_0$.

\section{Conclusion\label{conclusion}}

In this paper we present a new method of outlier rejection that does
not rely on priors and is capable of clipping misclassified first
overtone (FO) Cepheids from the fundamental mode (FM) Cepheid sample. The method
is similar to the outlier rejection method established by
E14. Both use the dispersion to correct the
underestimated errors from photometry. The difference is that our
median absolute deviation clipping method does not use an additional free
parameter.

We use the publicly available PHAT \citep{PHAT} data to obtain
near-infrared photometry of a subsample of Cepheids published in
K13. Our data reduction pipeline takes the HST and PS1
difference images into account in order to identify the correct source
in the PHAT data. With the MAD clipping method we
obtain a sample of 371 Cepheids with F110W and F160W photometry. The
sample consists of 319 FM, 16 FO and 36 type II Cepheids. 110 FM
Cepheids have periods of 10 days or more. The slopes of our PLRs for
Cepheids with periods of 10 or more days are shallower than the slopes
obtained by R12, but agree within the $1\sigma$ errors.

We check our sample for a broken slope in the PLR and find that a
broken slope describes the data significantly better than a linear slope.  

An estimation of the effect of our PLRs on the Hubble constant shows
that our sample gives a $3.2\%$ larger $H_0$ than the R12 sample.

With the full three years of PAndromeda data the Cepheid sample will
increase, especially toward longer periods. Additionally we will be
able to perform phase correction to the PHAT data. This will help to
distinguish between a broken slope PLR and a parabolic PLR. The phase
correction will also improve the dispersion further, resulting in an even
tighter constrained PLR.

\acknowledgments

 We are very grateful to the anonymous referee for the helpful
  comments.

We also want to thank Adam Riess for our fruitful discussion on crowding.

This research was supported by the DFG cluster of excellence Origin
and Structure of the Universe’ (www.universe-cluster.de).

Some of the data presented in this paper were obtained from the
Mikulski Archive for Space Telescopes (MAST). STScI is operated by the
Association of Universities for Research in Astronomy, Inc., under
NASA contract NAS5-26555. Support for MAST for non-HST data is
provided by the NASA Office of Space Science via grant NNX13AC07G and
by other grants and contracts.

WG gratefully acknowledges support for this work from the BASAL Centro
de Astrofisica y Tecnologias Afines (CATA) PFB-06/2007, and from the
Chilean Ministry of Economy, Development and Tourism's Millenium
Science Initiative through grant IC120009 awarded to the Millenium
Institute of Astrophysics (MAS). WG also very gratefully acknowledges
support from Prof. Ralf Bender for a research stay at the MPE Garching
and the University Observatory Munich.

This research was supported by the Munich Institute for Astro- and
Particle Physics (MIAPP) of the DFG cluster of excellence "'Origin and
Structure of the Universe''''.  
\clearpage
\appendix 

\section{Appendix}

Our sample will be published in electronic form on the CDS.

\subsection{Stampouts} 
\begin{figure}
\epsscale{1.0}
\plotone{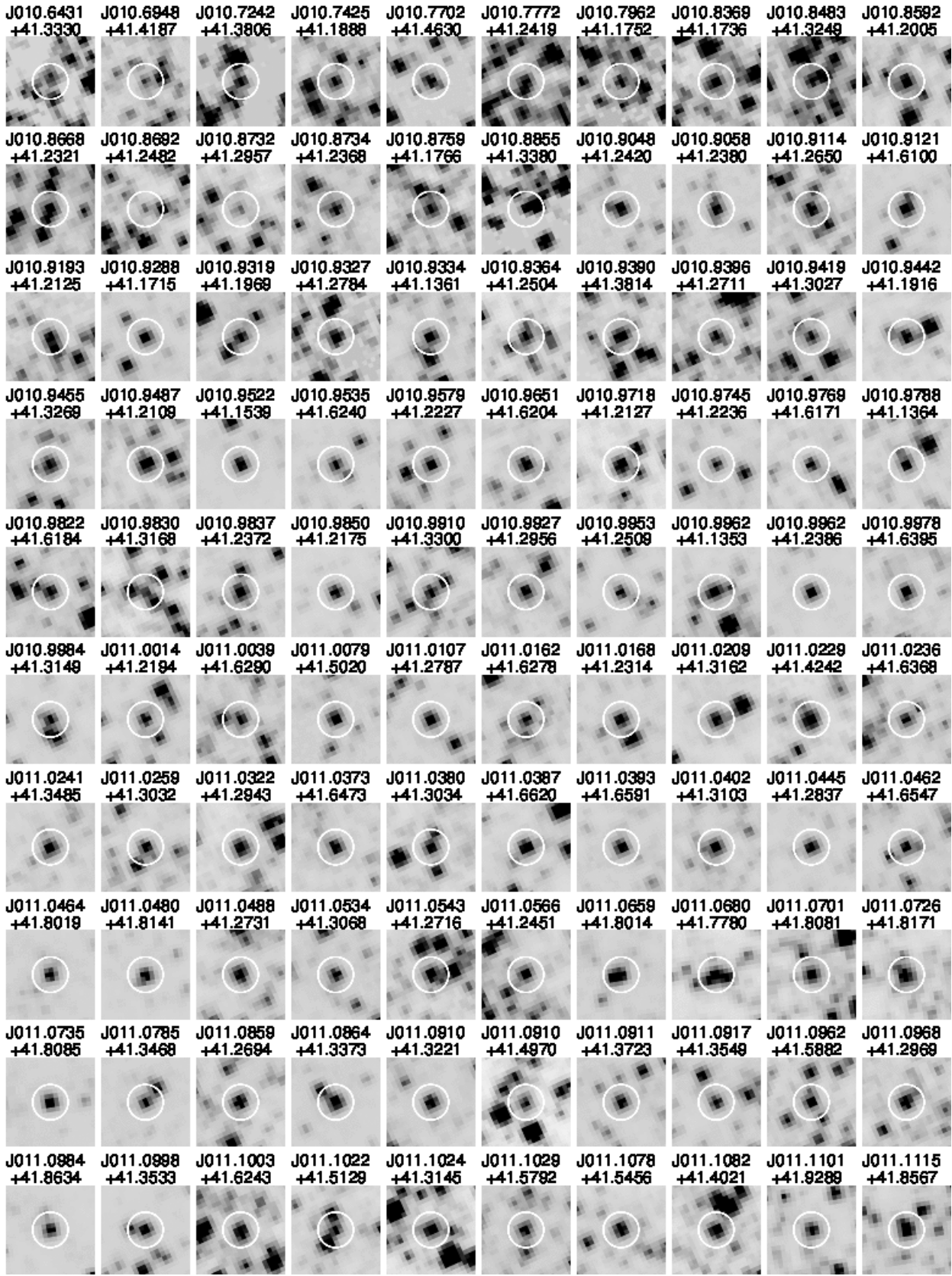}
\caption{Cepheid stampouts \label{fig_stamps_clean_1}}
\end{figure}

\begin{figure}
\epsscale{1.0}
\plotone{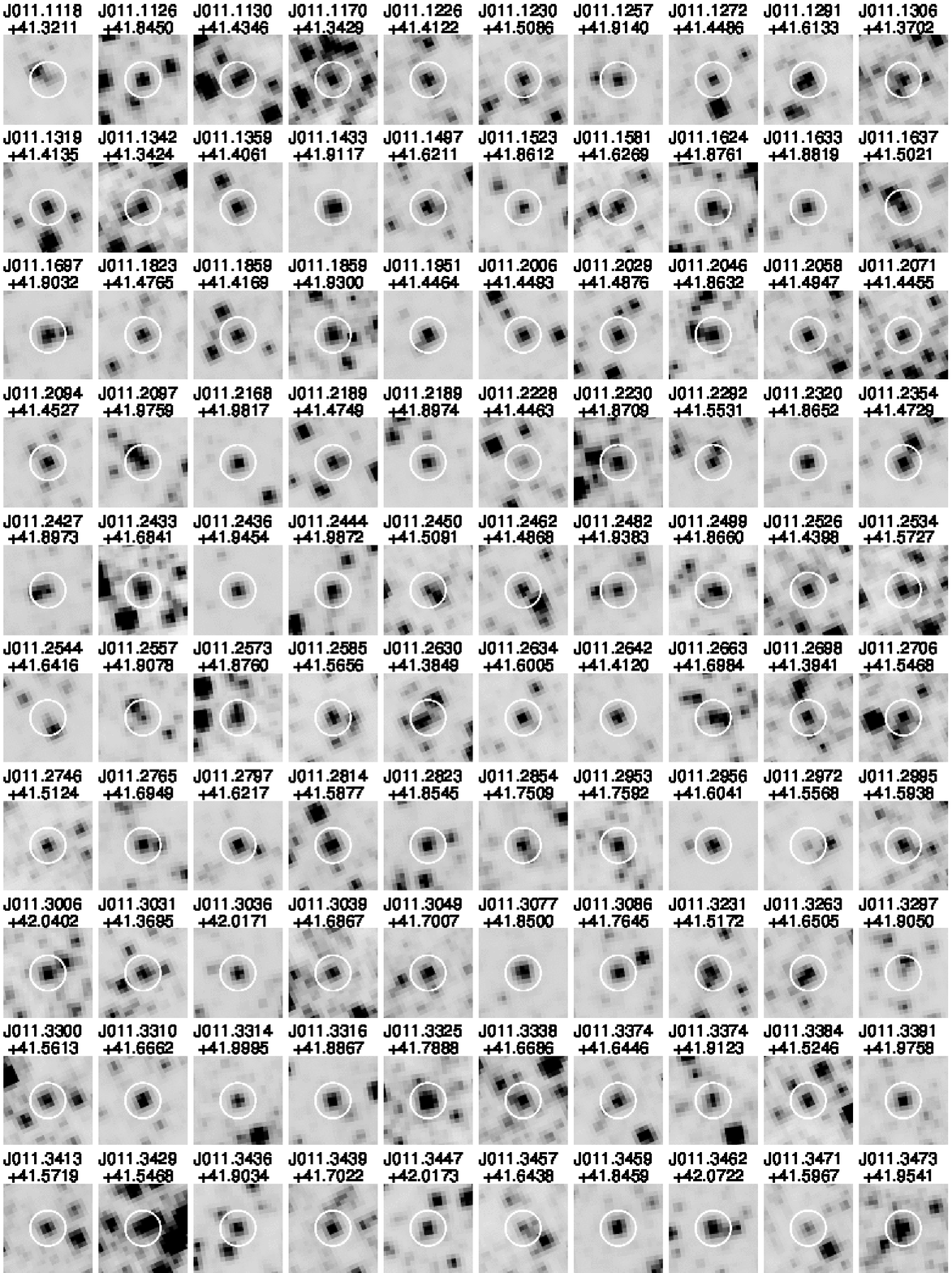}
\caption{Cepheid stampouts \label{fig_stamps_clean_2}}
\end{figure}

\begin{figure}
\epsscale{1.0}
\plotone{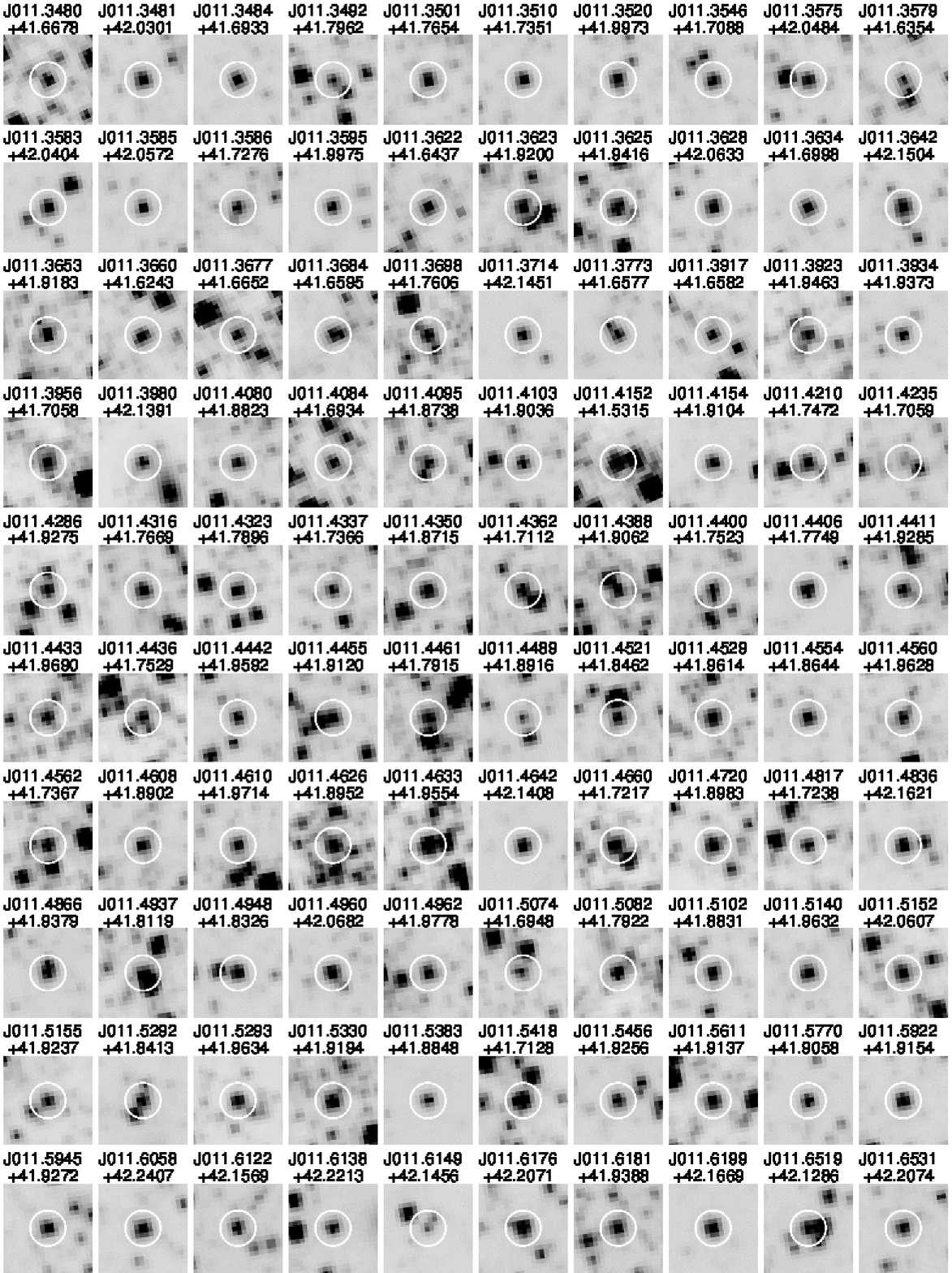}
\caption{Cepheid stampouts \label{fig_stamps_clean_3}}
\end{figure}

\begin{figure}
\epsscale{1.0}
\plotone{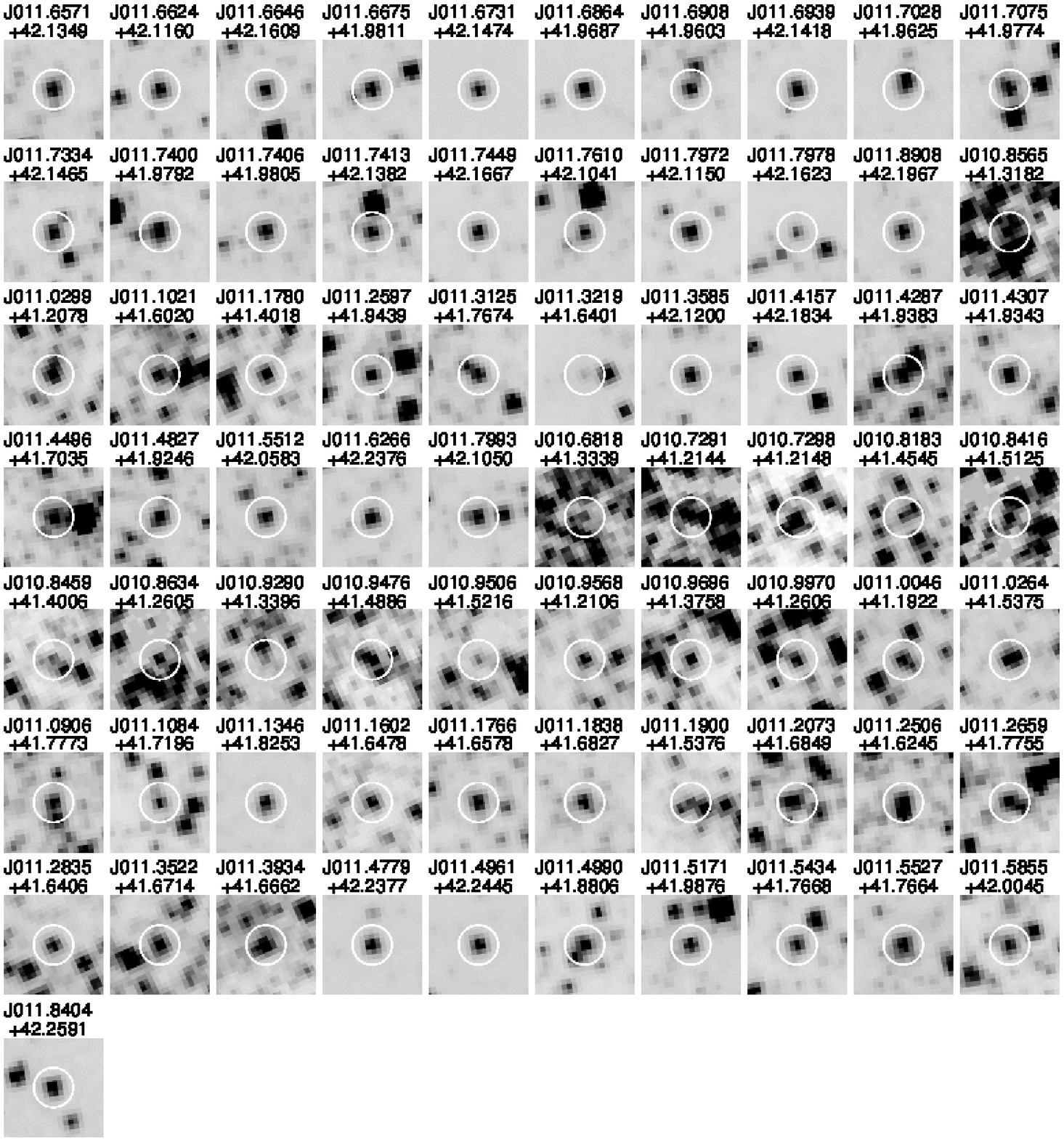}
\caption{Cepheid stampouts \label{fig_stamps_clean_4}}
\end{figure}

\begin{figure}
\epsscale{1.0}
\plotone{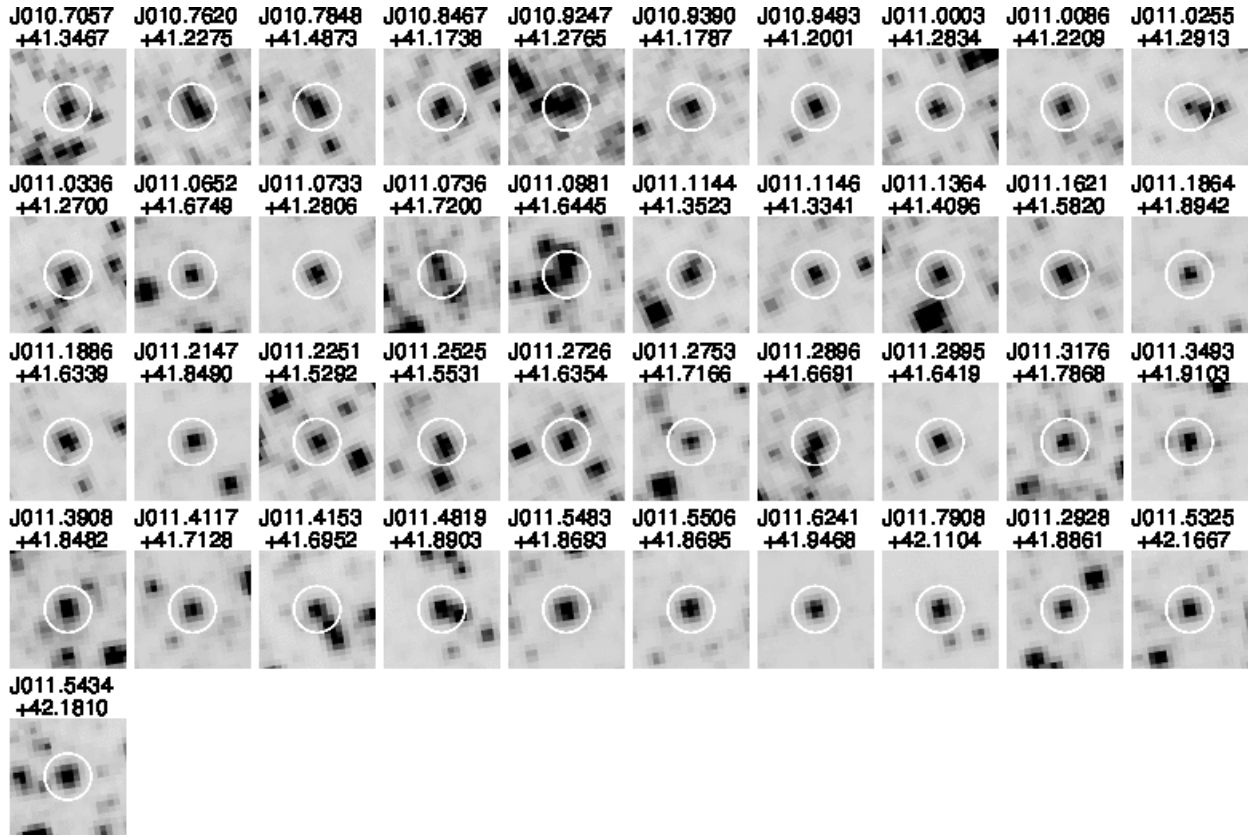}
\caption{Stampouts of the clipped Cepheids \label{fig_stamps_unclean_1}}
\end{figure}

The stampouts of the 371 Cepheids (319 FM Cepheids, 16 FO Cepheids and
36 type II Cepheids) can be seen in
Fig. \ref{fig_stamps_clean_1}, Fig. \ref{fig_stamps_clean_2},
Fig. \ref{fig_stamps_clean_3} and Fig. \ref{fig_stamps_clean_4}. The
stampouts for the clipped outliers are shown in
Fig. \ref{fig_stamps_unclean_1}. The scaling in each stampout is
different and calculated automatically. Therefore the
brightness between two stampouts cannot be compared. Each stampout
has the width of 2.5 arcsec and the white circle centered around the
Cepheid has a radius of 0.5 arcsec.

\clearpage

\subsection{Crowding test}

This section of the appendix provides Figures (Fig. \ref{fig_uncrowded_PLR_Wesenheit} - Fig. \ref{fig_uncrowded_method1_3}) and Tables (Table \ref{table_uncrowded_PLRs} - Table \ref{table_uncrowded_PLRs_parabola}) that include
only those Cepheids that have no source closer than 1.5 pixels
(c.f. Fig. \ref{fig_crowding}) i.e. are uncrowded. The uncrowded
sample consists of 265 Cepheids (217 FM Cepheids, 14 FO Cepheids and
34 type II Cepheids). For this sample 37 Cepheids were clipped (33 FM
Cepheids, 3 FO Cepheids and 1 type II Cepheid). The relevant F-test
values (c.f. section \ref{chapter_PLRs}) are
$F_{crit}=F(1,268;0.05)=3.88$, $F_{obs,Wesenheit}=8.08$,
$F_{obs,F110W}=12.13$ and $F_{obs,F160W}=13.48$. So the broken slopes
are still significant at a $2 \sigma$ level and the F110W and F160W
broken slopes are also still significant at a $3 \sigma$ level. Note
that the mean F160W offset in Fig. \ref{fig_compare_riess_160} changes
to $-0.018~\mathrm{mag}$ and the offset in
Fig. \ref{fig_compare_riess_110} changes to $-0.257~\mathrm{mag}$. 
Equation \ref{h0eqn3} changes to 
\begin{equation}
\Delta M  = < m(\log(P)=1.2)_{\mathrm{R12}} - m(\log(P)=1.2)_{\mathrm{K14}} > =
17.689~\mathrm{mag} - 17.772~\mathrm{mag} = -0.083~\mathrm{mag}
\end{equation}
which implies for Equation \ref{h0eqn4}:
\begin{equation}
\Delta \mu = (m-M)_{\mathrm{R12}} - (m-M)_{\mathrm{K14}} = \Delta m - \Delta M = -
\Delta M = 0.083~\mathrm{mag}
\end{equation}
and therefore
\begin{equation}
H_{\mathrm{0,K14}} = 10^{[\Delta \mu_0 / 5]} \cdot H_{\mathrm{0,R12}} = 1.039 \cdot H_{\mathrm{0,R12}}
\end{equation}
So the $3.2\%$ increase to $3.9\%$.

\begin{figure}
\epsscale{1.0}
\plotone{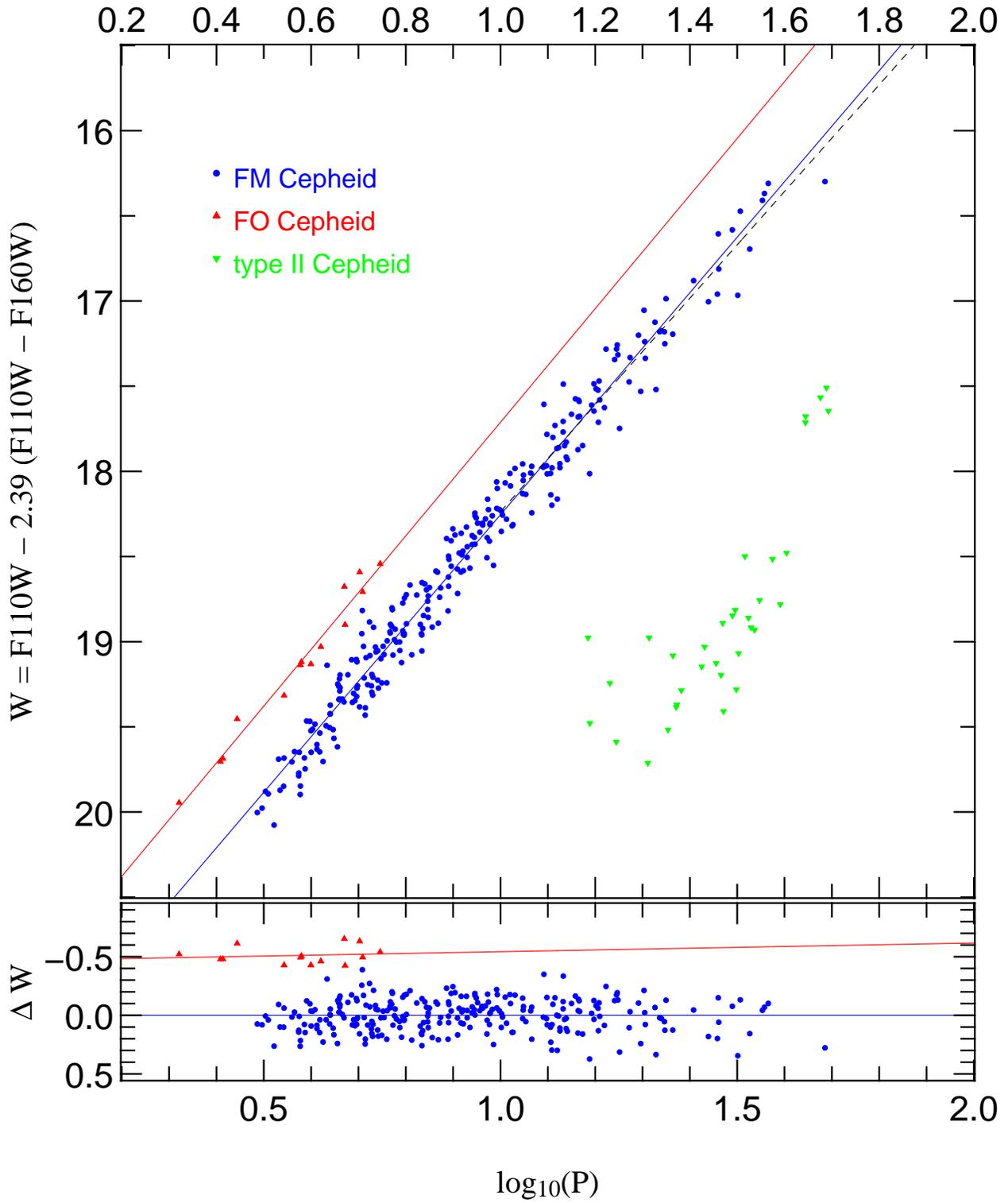}
\caption{Same as Fig. \ref{fig_PLR_Wesenheit} but only for uncrowded
  Cepheids (217 FM Cepheids, 14 FO Cepheids and 34 type II
  Cepheids). \label{fig_uncrowded_PLR_Wesenheit}}
\end{figure}

\begin{figure}
\epsscale{1.0}
\plotone{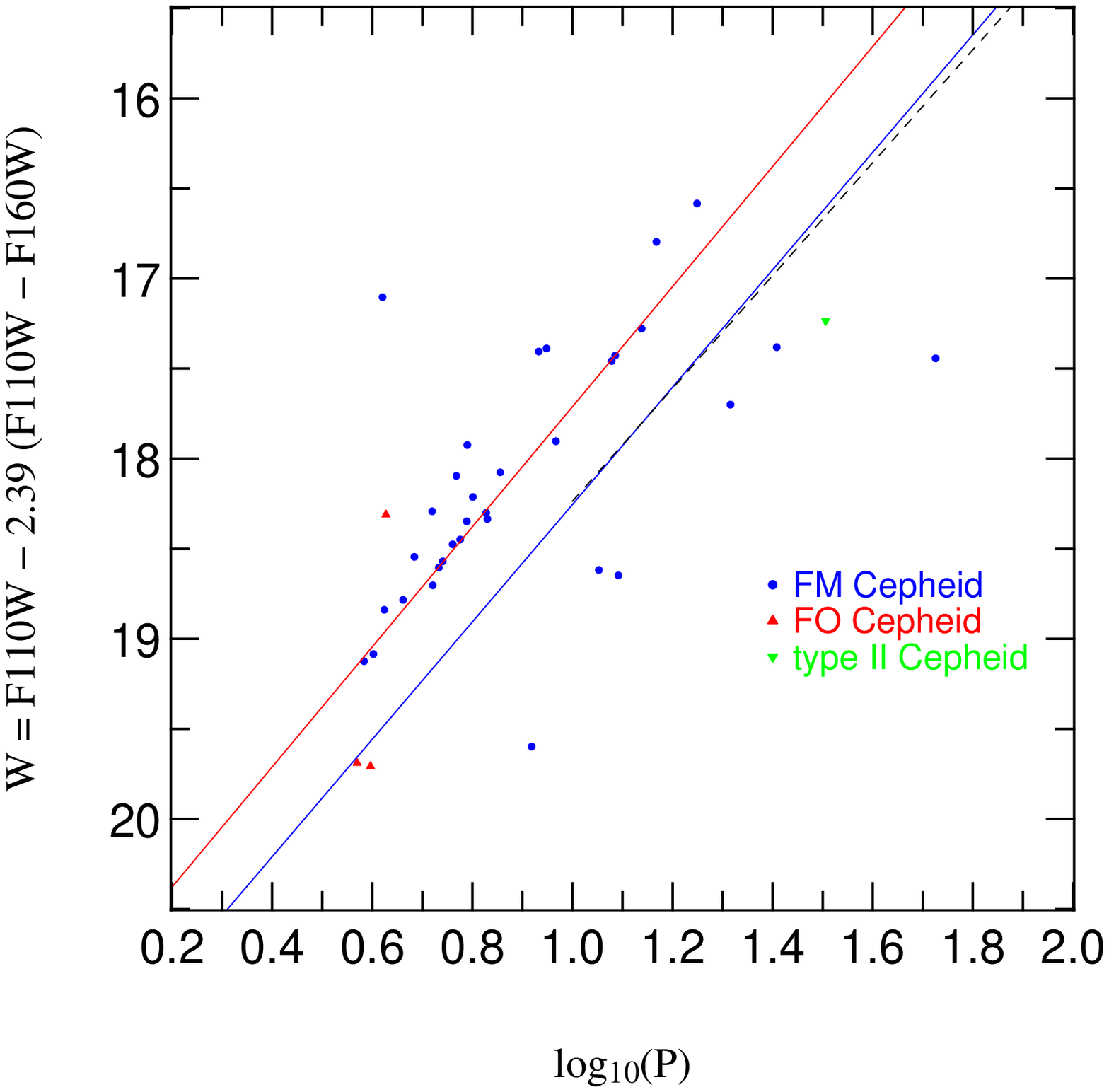}
\caption{Same as Fig. \ref{fig_PLR_Wesenheit_outliers}, but only for uncrowded Cepheids. There
  are 37 clipped Cepheids (33 FM Cepheids, 3 FO Cepheids and 1 type II Cepheid).\label{fig_uncrowded_PLR_Wesenheit_outliers}}
\end{figure}

\begin{deluxetable}{cccccccccc}
\tabletypesize{\scriptsize}
\rotate
\tablecaption{PLR fit parameters\label{table_uncrowded_PLRs}}
\tablewidth{0pt}
\tablehead{
\colhead{$\#$} & \colhead{band} & \colhead{type} & \colhead{range} & \colhead{$N_{fit}$} &
\colhead{a (log P = 1)} & \colhead{slope b} & \colhead{$\sigma$} &
\colhead{$\sigma_{int}$\tablenotemark{a}} & \colhead{$\chi_{d.o.f.}^2$\tablenotemark{b}}
}
\startdata
 1 & F110W & FM & all & 271 & 19.515 ( 0.007) & -2.778 ( 0.032) &   0.209 & - &  1.000 \\
 2 & F110W & FM & log P $>$ 1 & 93 & 19.464 ( 0.055) & -2.483 ( 0.155) &   0.251 & - &  1.435 \\
 3 & F110W & FO & all & 14 & 18.947 ( 0.070) & -2.714 ( 0.152) &   0.112 & - &  1.000 \\
 4 & F160W & FM & all & 271 & 18.987 ( 0.002) & -2.979 ( 0.023) &   0.158 & - &  1.000 \\
 5 & F160W & FM & log P $>$ 1 & 93 & 18.950 ( 0.023) & -2.755 ( 0.125) &   0.184 & - &  1.348 \\
 6 & F160W & FO & all & 14 & 18.429 ( 0.072) & -2.973 ( 0.154) &   0.087 & - &  1.000 \\
 7 & Wesenheit & FM & all & 271 & 18.255 ( 0.005) & -3.259 ( 0.080) &   0.137 & - &  1.000 \\
 8 & Wesenheit & FM & log P $>$ 1 & 93 & 18.236 ( 0.017) & -3.132 ( 0.086) &   0.150 & - &  1.209 \\
 9 & Wesenheit & FO & all & 14 & 17.712 ( 0.100) & -3.333 ( 0.176) &   0.077 & - &  1.000 \\
\enddata
\tablenotetext{a}{internal scatter as defined by E14}
\tablenotetext{b}{reduced $\chi^2$}
\tablecomments{Same as Table \ref{table_PLRs} but for uncrowded Cepheids.}
\end{deluxetable}

\begin{deluxetable}{ccccccc}
  \tabletypesize{\scriptsize} \rotate \tablecaption{broken slope PLR
    fit parameters\label{table_uncrowded_PLRs_broken}} \tablewidth{0pt} \tablehead{
    \colhead{$\#$} & \colhead{band} & \colhead{$b_{\log(P)\leq1}$} &
    \colhead{$b_{\log(P)>1}$} &\colhead{$a_{\log(P)=1}$} &
    \colhead{$\sigma$} & \colhead{$\chi_{d.o.f.}^2$\tablenotemark{a}}
}
\startdata
1 & Wesenheit & -3.411 ( 0.058) & -3.077 ( 0.080) & 18.219 ( 0.017) &  0.135 &  0.974 \\
2 & F110W & -3.071 ( 0.079) & -2.430 ( 0.136) & 19.446 ( 0.020) &  0.205 &  0.958 \\
3 & F160W & -3.213 ( 0.061) & -2.701 ( 0.103) & 18.932 ( 0.015) &  0.155 &  0.952 \\
\enddata
\tablenotetext{a}{reduced $\chi^2$}
\tablecomments{Same as Table \ref{table_PLRs_broken} but for uncrowded Cepheids.}
\end{deluxetable}

\begin{deluxetable}{ccccccc}
  \tabletypesize{\scriptsize} \rotate \tablecaption{parabola PLR
    fit parameters\label{table_uncrowded_PLRs_parabola}} \tablewidth{0pt} \tablehead{
    \colhead{$\#$} & \colhead{band} & \colhead{$a_{\log(P)=1}$} &
    \colhead{$b$} &\colhead{$c$} &
    \colhead{$\sigma$} & \colhead{$\chi_{d.o.f.}^2$\tablenotemark{a}}
}
\startdata
1 & Wesenheit & 18.233 ( 0.011) & -3.253 ( 0.030) &  0.318 ( 0.121) &  0.135 &  0.973 \\
2 & F110W & 19.475 ( 0.017) & -2.767 ( 0.047) &  0.595 ( 0.186) &  0.205 &  0.958 \\
3 & F160W & 18.955 ( 0.012) & -2.970 ( 0.036) &  0.479 ( 0.142) &  0.155 &  0.952 \\
\enddata
\tablenotetext{a}{reduced $\chi^2$}
\tablecomments{Same as Table \ref{table_PLRs_parabola} but for uncrowded Cepheids.}
\end{deluxetable}

\begin{figure}
\epsscale{1.0}
\plotone{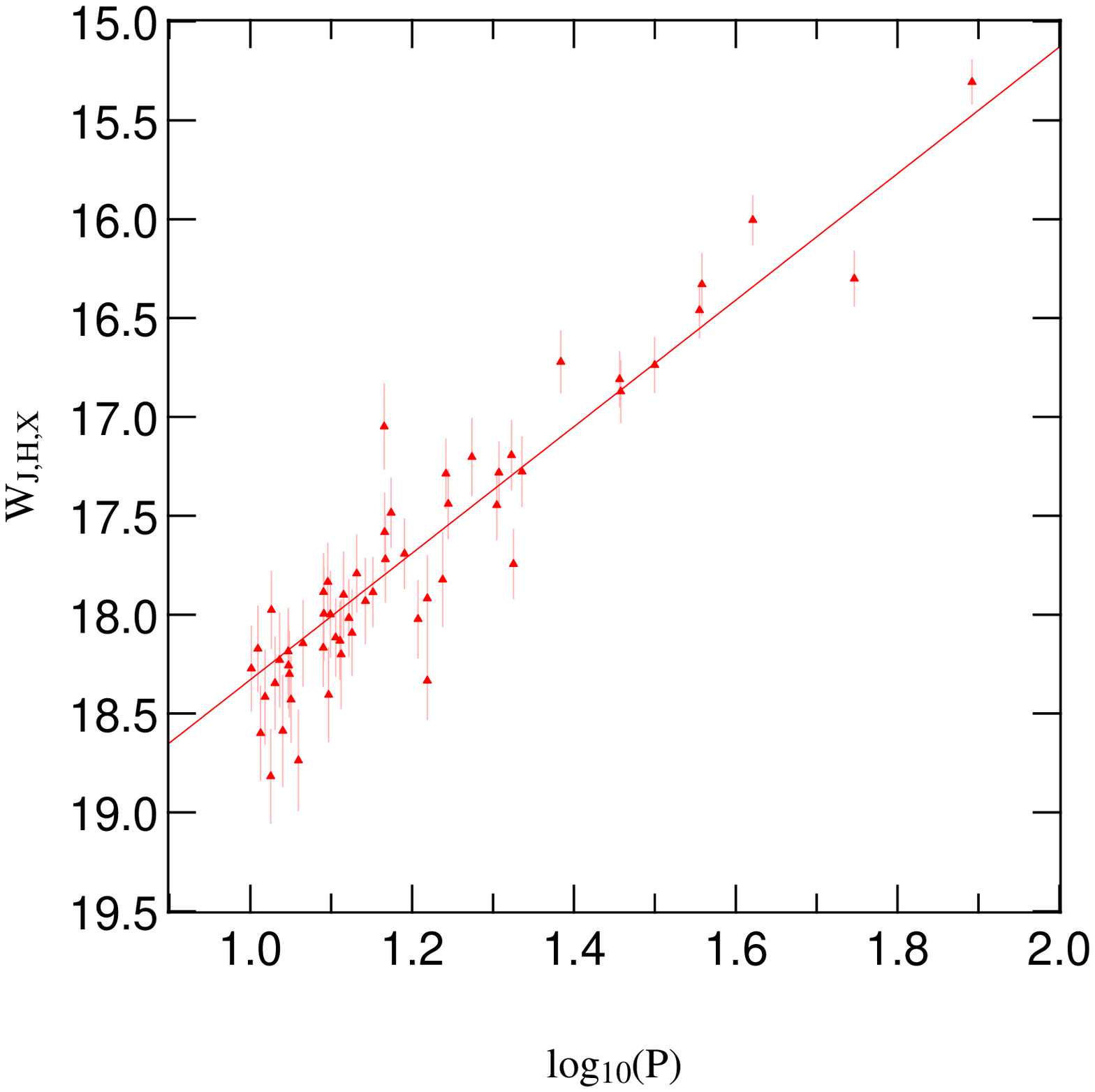}
\caption{Same as Fig. \ref{fig_method0_3} but for 56 uncrowded Cepheids. \label{fig_uncrowded_method0_3}}
\end{figure}

\begin{figure}
\epsscale{1.0}
\plotone{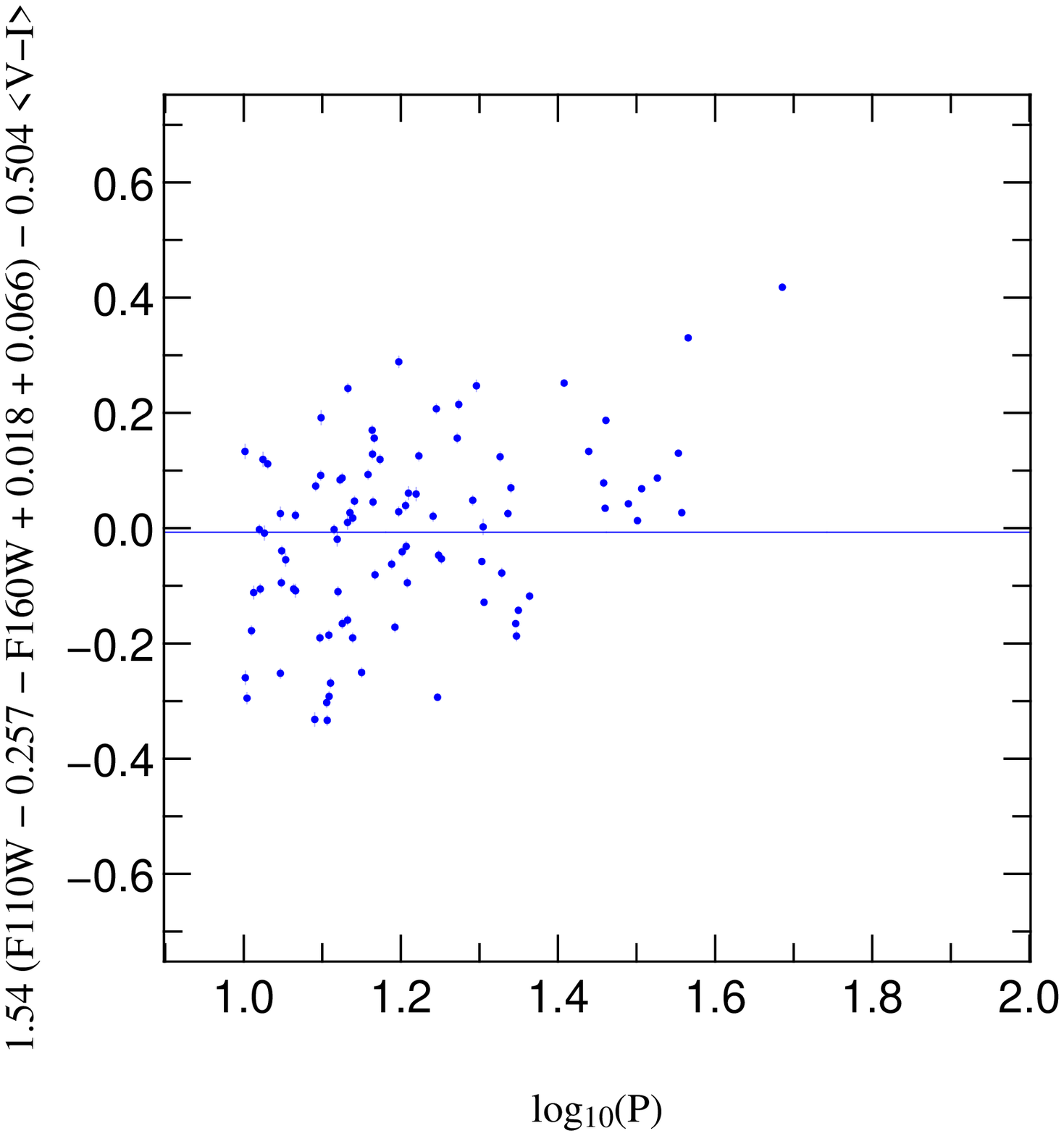}
\caption{Same as Fig. \ref{fig_method1_2} but for uncrowded Cepheids. \label{fig_uncrowded_method1_2}}
\end{figure}

\begin{figure}
\epsscale{1.0}
\plotone {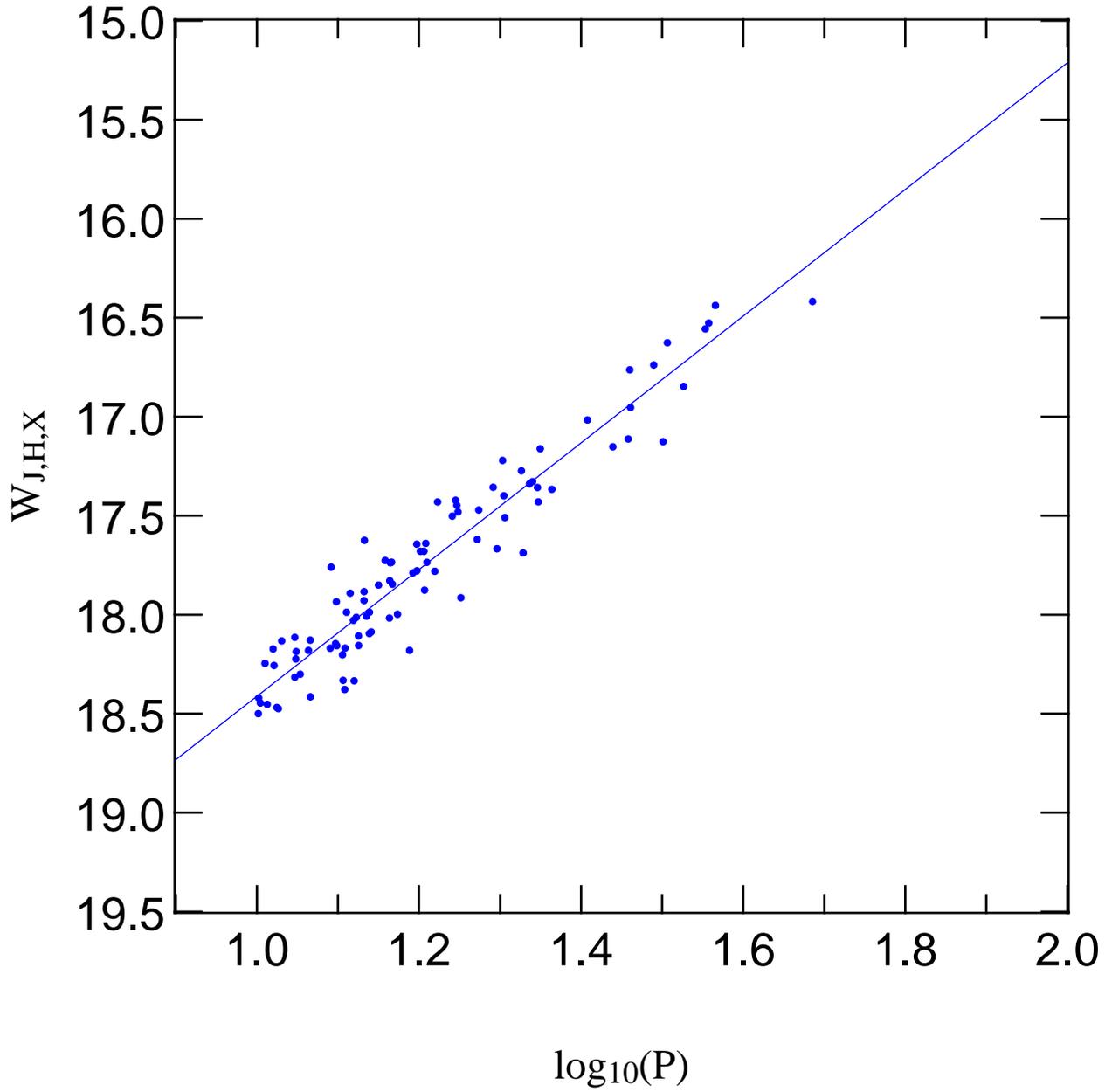}
\caption{Same as Fig. \ref{fig_method1_3} but for uncrowded Cepheids. \label{fig_uncrowded_method1_3}}
\end{figure}
 
\clearpage

\subsection{Suspension point}

Due to the fact that our data are random phased we
cannot be sure about the correct suspension point. We believe it is
better to determine the suspension point with phase corrected data.
For the interested reader we provide Table
\ref{table_PLRs_broken_appendix} that shows the fit parameters for
different suspension points. The best fit for the broken slope is
around 8 to 9 days but the observed F-ratio is above the critical F-value
for all fits, although also this value favors a suspension point
around 8 to 9 days. 

\begin{deluxetable}{cccccccc}
  \tabletypesize{\scriptsize} \rotate \tablecaption{broken slope PLR
    fit parameters\label{table_PLRs_broken_appendix}} \tablewidth{0pt} \tablehead{
    \colhead{$x_0$} & \colhead{band} & \colhead{$b_{\log(P)\leq1}$} &
    \colhead{$b_{\log(P)>1}$} &\colhead{$a_{\log(P)=1}$} &
    \colhead{$\sigma$} & \colhead{$\chi_{d.o.f.}^2$\tablenotemark{a}} &\colhead{$F_{\mathrm{obs}}$}
}
\startdata
  5.000 & Wesenheit & -3.813 ( 0.138) & -3.209 ( 0.033) & 19.214 ( 0.012) &  0.136 &  0.978 &  8.195\\
  5.000 & F110W & -3.366 ( 0.218) & -2.684 ( 0.056) & 20.322 ( 0.020) &  0.203 &  0.988 &  4.707 \\
  5.000 & F160W & -3.553 ( 0.166) & -2.904 ( 0.042) & 19.858 ( 0.015) &  0.154 &  0.980 &  7.461 \\
  6.000 & Wesenheit & -3.598 ( 0.090) & -3.189 ( 0.040) & 18.952 ( 0.014) &  0.136 &  0.979 &  7.845\\
  6.000 & F110W & -3.240 ( 0.133) & -2.634 ( 0.064) & 20.090 ( 0.020) &  0.202 &  0.979 &  7.841 \\
  6.000 & F160W & -3.390 ( 0.087) & -2.866 ( 0.048) & 19.613 ( 0.014) &  0.153 &  0.972 & 10.221 \\
  7.000 & Wesenheit & -3.508 ( 0.068) & -3.170 ( 0.043) & 18.731 ( 0.013) &  0.136 &  0.979 &  7.710\\
  7.000 & F110W & -3.202 ( 0.097) & -2.567 ( 0.072) & 19.891 ( 0.020) &  0.201 &  0.965 & 12.573 \\
  7.000 & F160W & -3.330 ( 0.068) & -2.819 ( 0.055) & 19.405 ( 0.014) &  0.152 &  0.960 & 14.168 \\
  8.000 & Wesenheit & -3.474 ( 0.056) & -3.141 ( 0.045) & 18.538 ( 0.014) &  0.136 &  0.976 &  8.923\\
  8.000 & F110W & -3.154 ( 0.077) & -2.503 ( 0.079) & 19.721 ( 0.020) &  0.200 &  0.955 & 15.840 \\
  8.000 & F160W & -3.288 ( 0.057) & -2.770 ( 0.061) & 19.225 ( 0.013)  &  0.151 &  0.951 & 17.470 \\
  8.500 & Wesenheit & -3.455 ( 0.054) & -3.130 ( 0.047) & 18.451 ( 0.014) &  0.136 &  0.976 &  8.959\\
  8.500 & F110W & -3.117 ( 0.084) & -2.483 ( 0.085) & 19.648 ( 0.020) &  0.200 &  0.955 & 15.773 \\
  8.500 & F160W & -3.259 ( 0.050) & -2.754 ( 0.062) & 19.147 ( 0.013) &  0.151 &  0.951 & 17.435 \\
  9.000 & Wesenheit & -3.440 ( 0.045) & -3.119 ( 0.050) & 18.370 ( 0.014) &  0.136 &  0.975 &  8.969\\
  9.000 & F110W & -3.084 ( 0.079) & -2.464 ( 0.090) & 19.579 ( 0.020) &  0.200 &  0.956 & 15.456 \\
  9.000 & F160W & -3.233 ( 0.051) & -2.739 ( 0.068) & 19.073 ( 0.013) &  0.151 &  0.951 & 17.183 \\
 10.000 & Wesenheit & -3.411 ( 0.038) & -3.103 ( 0.060) & 18.221 ( 0.013) &  0.136 &  0.978 &  8.237\\
 10.000 & F110W & -3.028 ( 0.078) & -2.433 ( 0.105) & 19.455 ( 0.021) &  0.200 &  0.960 & 14.121 \\
 10.000 & F160W & -3.188 ( 0.050) & -2.714 ( 0.069) & 18.938 ( 0.014) &  0.152 &  0.956 & 15.707 \\
 10.470 & Wesenheit & -3.399 ( 0.039) & -3.097 ( 0.063) & 18.156 ( 0.014) &  0.136 &  0.979 &  7.723\\
 10.470 & F110W & -3.007 ( 0.076) & -2.419 ( 0.112) & 19.401 ( 0.022) &  0.201 &  0.962 & 13.512 \\
 10.470 & F160W & -3.171 ( 0.048) & -2.703 ( 0.072) & 18.880 ( 0.015) &  0.152 &  0.958 & 14.939 \\
 11.000 & Wesenheit & -3.386 ( 0.039) & -3.092 ( 0.067) & 18.087 ( 0.015) &  0.136 &  0.981 &  7.140\\
 11.000 & F110W & -2.987 ( 0.072) & -2.401 ( 0.116) & 19.343 ( 0.022) &  0.201 &  0.963 & 13.035 \\
 11.000 & F160W & -3.154 ( 0.047) & -2.691 ( 0.076) & 18.817 ( 0.015) &  0.152 &  0.960 & 14.240 \\
 12.000 & Wesenheit & -3.365 ( 0.040) & -3.086 ( 0.083) & 17.967 ( 0.015) &  0.137 &  0.985 &  5.971\\
 12.000 & F110W & -2.956 ( 0.070) & -2.367 ( 0.119) & 19.241 ( 0.022) &  0.201 &  0.965 & 12.331 \\
 12.000 & F160W & -3.127 ( 0.040) & -2.668 ( 0.091) & 18.707 ( 0.017) &  0.152 &  0.963 & 13.031 \\
 15.000 & Wesenheit & -3.325 ( 0.037) & -3.066 ( 0.118) & 17.658 ( 0.017) &  0.137 &  0.991 &  3.977\\
 15.000 & F110W & -2.873 ( 0.057) & -2.325 ( 0.190) & 18.990 ( 0.026) &  0.202 &  0.978 &  8.162 \\
 15.000 & F160W & -3.063 ( 0.044) & -2.636 ( 0.139) & 18.432 ( 0.020) &  0.153 &  0.977 &  8.618 \\
\enddata
\tablenotetext{a}{reduced $\chi^2$}
\tablecomments{Same as Fig. \ref{table_PLRs_broken} but for
    different suspension points $x_0$. Additionally the corresponding
    $F_{\mathrm{obs}}$ (c.f. section \ref{chapter_PLRs}) is given.}
\end{deluxetable}

\end{document}